\documentclass[fleqn,usenatbib]{mnras}

\usepackage{newtxtext,newtxmath}

\usepackage[T1]{fontenc}
\usepackage{ae,aecompl}

\usepackage{graphicx}	% Including figure files
\usepackage{amsmath}	% Advanced maths commands
\usepackage{natbib}     % Bibliography
\usepackage{longtable}  % For the SN table
\usepackage{soul}

\newcommand{\dd}{{\rm d}}

\newcommand{\change}{}

\title[DTD of SNe from IFS]{The delay time distribution of supernovae from integral-field spectroscopy of nearby galaxies}

\author[Asier Castrillo et al.]{Asier Castrillo$^{1,2}$\thanks{E-mail: asier.castrillo@uam.es}, 
Yago Ascasibar$^{1,2}$,
Llu\'is Galbany$^{3}$,
Sebasti\'an F. S\'anchez$^{4}$,
\newauthor
Carles Badenes$^{5}$,
Joseph P. Anderson$^{6}$,
Hanindyo Kuncarayakti$^{7,8}$,
Joseph D. Lyman$^{9}$,
\newauthor
Angeles I. D\'iaz$^{1,2}$ \\
$^{1}$Departamento de F\'isica Te\'orica, Universidad Aut\'onoma de Madrid, E-28049, Cantoblanco (Madrid), Spain.\\
$^{2}$Research Center of Advanced Fundamental Physics (CIAFF-UAM), Universidad Aut\'onoma de Madrid, E-28049, Cantoblanco (Madrid), Spain.\\
$^{3}$Departamento de F\'isica Te\'orica y del Cosmos, Universidad de Granada, E-18071 Granada, Spain.\\
$^{4}$Instituto de Astronom\'ia, Universidad Nacional Aut\'onoma de M\'exico, A.P. 70-264, 04510 M\'exico, D.F., Mexico.\\
$^{5}$PITT PACC, Department of Physics and Astronomy, University of Pittsburgh, Pittsburgh, PA 15260, USA.\\
$^{6}$European Southern Observatory, Alonso de C\'ordova 3107 Casilla 19001, Vitacura, Santiago, Chile.\\
$^{7}$Department of Physics and Astronomy, University of Turku, FI−20014 Turku, Finland.\\
$^{8}$Finnish Centre for Astronomy with ESO (FINCA), FI-20014 University of Turku, Finland.\\
$^{9}$Department of Physics, University of Warwick, Coventry, CV4 7AL, UK.
}
\date{Received date: 9 November 2020 ; accepted date: 7 December 2020}

\pubyear{2020}

\DeclareUnicodeCharacter{2212}{-}
\begin{document}
\label{firstpage}
\pagerange{\pageref{firstpage}--\pageref{lastpage}}
\maketitle

% Abstract of the paper
\begin{abstract}
Constraining the delay-time distribution (DTD) of different supernova (SN) types can shed light on the timescales of galaxy chemical enrichment and feedback processes affecting galaxy dynamics, and SN progenitor properties. Here, we present an approach to recover SN DTDs based on integral field spectroscopy (IFS) of their host galaxies. Using a statistical analysis of a sample of 116 supernovae in 102 galaxies, we evaluate different DTD models for SN types Ia (73), II (28) and Ib/c (15). We find the best SN Ia DTD fit to be a power law with an exponent $\alpha = -1.1\pm 0.3$ (50\% confidence interval), and a time delay (between star formation and the first SNe) $\Delta = 50^{+100}_{-35}~Myr$ (50\% C.I.). For core collapse (CC) SNe,\change{ both of the Zapartas et al. (2017) DTD models for single and binary stellar evolution are consistent with our results.} For SNe II and Ib/c, we find a correlation with a Gaussian DTD model with $\sigma = 82^{+129}_{-23}~Myr$ and $\sigma = 56^{+141}_{-9}~Myr$ (50\% C.I.) respectively. This analysis demonstrates that integral field spectroscopy opens a new way of studying SN DTD models in the local universe.
\end{abstract}


%%%%%%%%%%%%%%%%% BODY OF PAPER %%%%%%%%%%%%%%%%%%

%-----------------------------------------------------------------------------------
\section{Introduction}
%-----------------------------------------------------------------------------------

Supernovae (SNe) represent an important source of stellar feedback and chemical enrichment of the interstellar medium. Despite their importance, the physical mechanisms that trigger the explosions, and the relation between SNe and their progenitors is not fully understood. Historically, SNe were first classified based on the presence (type II) or lack (type I) of Hydrogen features in their observed spectra \citep{Minkowski1941}. Although most type I SNe show a very prominent Si II feature (now labelled type Ia), some objects do not, and they were later divided into new subtypes by the presence (type Ib) or absence (type Ic) of Helium features.

Currently, it is commonly accepted that SNe II, Ib and Ic result from the core collapse (CC) of massive stars \citep{CoreCollapse1984}, whereas SNe Ia are produced by the thermonuclear explosion of Carbon-Oxygen white dwarfs (WD) in binary systems \citep{Maoz2014,WangHan2012,IaExplosion2000}.
The precise nature of the WD companion is still an active area of research \citep{Howell2011,Maoz2008,Wong2019}.
In single-degenerate (SD) scenarios, the mass of the WD increases due to accretion from a non-degenerate stellar companion, such as a main sequence star \citep{Whelan1973} or a He star donor \citep{Yoon2003}.
When the WD mass approaches the Chandrasekhar mass limit, the central regions of the WD cross the C ignition curve and explode in a thermonuclear runaway.
In the double-degenerate (DD) scenario \citep{Webbink1984}, two WDs merge after losing angular momentum.
The resulting object approaches the Chandrasekhar mass and explodes.
However, recent studies \citep{Pakmor2010} have shown that sub-Chandrasekhar SN explosions can occur through a double detonation process.
This motivates a new series of scenarios \citep[e.g.][]{Fink2007, Guillochon2010, Shen2018} where a degenerate companion (although a similar process is possible in the SD case) transfers mass to the WD, the He envelope explodes, and this triggers the explosion of the C/O core.

Due to the different time scales of the processes that govern the late evolutionary stages of SN progenitors, different SN types are expected to appear after a star formation event according to a certain delay-time distribution (DTD) that connects the star formation rate with the SN rate (SNR) of each type (Ia, II, Ib/c; \citealt{Yungelson2000}).
More precisely, the DTD is the hypothetical SNR that would follow an instantaneous burst of star formation (the expected number of SNe as a function of time per unit stellar mass formed). This way, the SNR $R \,(t)$ in a certain galaxy (or region) at a given time $t$ is given by the convolution of the DTD with the star formation history (SFH) $\Psi(t)$:
\begin{equation}
	R(t) = \int_0^t {\rm DTD}(\tau)\ \Psi(t-\tau)\ {\rm d}\tau
\end{equation}

The initial mass function (IMF) and the stellar lifetimes of the population being considered are the theoretical main ingredients that are required to estimate the DTD of CC SNe. But in practice the DTD is derived from a SN catalogue and a collection of SFHs from the stellar populations that spawned the SNe (either from individual galaxies or from a cosmological volume).
The exact form of the IMF, the statistical distribution of the number of stars that are born with a given mass, is still debated \citep{Salpeter1955, Kroupa2007}. However, there is an overall agreement in that most stars over approximately 8 $M_{\odot}$ end their lives as CC SNe. Recent studies have found that some stars more massive than 20 $M_{\odot}$ may not produce SNe \citep{Sukhbold2016}, but this mass range represents a relatively small fraction ($\sim 10$\% in mass, $<1$\% in number) of the total stellar population.
The most massive stars have the shortest lifetimes, but it is expected that all CC SNe associated with a given star formation burst should explode within a time interval between 3 and 50 Myr, assuming single star progenitors models and some metallicity and model dependence \citep{Georgy2013, zapartas2017}.

As it is common to find young massive stars in close binary systems \citep{Sana2012}, recent studies \citep[e.g.][]{DeDonder2003,Eldridge2015,Schady2019,Zapartas2019} \change{have explored the impact of CCSNe on the DTD.}
We expect that these massive stars interact with their companions, transferring mass through Roche-lobe overflow \citep{Poelarends2017,Kiewicz2019,Naiman2020}, common-envelope phases \citep{Lohev2019, Grichener2019}, or mergers.
Such interaction can significantly affect the evolution of both stars and the possible development of a CC SN \citep{Eldridge2008, Yoon2010}, implying that there is a population of stars less massive than 8~M$_\odot$ that, due to mass exchange in binary systems, may terminate their lives as CC SNe.
This population, with a characteristic time \change{delay between 50 and 200 Myr}, is usually referred to as late CC SNe. They are expected to represent around 15\% of the total number of CC SNe \citep{zapartas2017}.

The SN Ia DTD is more difficult to constrain from the theoretical perspective and also for the observational part. It is more difficult to derive the progenitor cluster of SN Ia; they disperse on galactic mixing timescales that are shorter than SN Ia timescales \citep{Clusters2003, MaozMannucci2012}, \change{You can avoid this problem taking into account the SFH from the entire galaxy}.
White dwarfs are the remnants of stars between 0.5 and 8 solar masses, which have much longer lifetimes.
Stars near the massive end of that range generate the first WDs relatively early, but they are outnumbered by the less massive ones that make the bulk of the population \citep{Nelemans2001}.
Since SNe Ia take place in close binary systems, \change{we need to know the formation rate} \citep[this a much debated topic][]{MoeDiStefano2017, Moe2018, Moe2019}, as well as the characteristic times of the different mass transport mechanisms.
In the simplest case \citep[the WD-WD merger scenario;][]{Ruiter2011}, it is also important to know the initial distance between the two objects and how fast their orbit will decay until the finally merge.
\change{The orbital decay process is driven by angular momentum loss due to gravitational wave emission}, yielding a DTD with a smooth tail at long times that can be approximated as a power law.
\change{Similar results have also been reported \citep[e.g.][]{Mennekens2010,Toonen2012,Bours2013} for the SD and the DD scenarios using binary stars population synthesis models.}

Observational studies first suggested a bimodal population of `prompt' and `tardy' SNe Ia \citep{Strolger2004, Mannucci2006}. 
However, when large galaxy-survey monitoring campaigns became available, \citet{MaozMannucci2012DTD} found a SNIa DTD compatible with a smooth function with a long tail:
\begin{equation}
	\frac{{\rm DTD_{Ia}}(\tau)}{\rm yr^{-1}\ M_{\odot}^{-1}}
	= 4 \times 10^{-13}	\left( \frac{t}{\rm Gyr}\right)^{-1} 
\end{equation}
which also provides a good fit to the observed [$\alpha$/Fe] element abundance ratio \citep{Walcher2016}.
Recent studies \citep[e.g.][]{Friedmann2018, Heringer2017, Heringer2019} find different values for the exponent of the DTD power law, the normalisation, and the time delay until the first explosion of a SN Ia (the turn-on time), as well as possible differences between field and cluster galaxies \citep{Maoz2017}.

In the present work, we take advantage of the reconstruction of the resolved SFH of galaxies spaxel by spaxel made possible by integral-field spectroscopy (IFS).
Section~\ref{sec_data} describes our observational data. Data analysis is fully described in Section~\ref{sec_SNrate}, devoted to the estimation of the SNR, and Section~\ref{subsec_statistic}, discussing the statistical foundations of our methodology.
%We evaluate whether specific forms of the DTD are statistically compatible with the observed distribution of SNe in our galaxy sample, providing additional independent constrains on the DTD of different supernova types.
Section~\ref{sec_results} presents the constraints on the DTD of Ia, II, and Ib/c SNe in the local Universe, and our main results and conclusions are discussed in Section~\ref{sec_discusion}.

%-----------------------------------------------------------------------------------
\section{Data Sample}
\label{sec_data}
%-----------------------------------------------------------------------------------

All the data used in the present work come from observations by the Multi Unit Spectroscopic Explorer (MUSE; \citealt{2010SPIE.7735E..08B}), an integral-field spectrograph mounted on the VLT UT4 telescope at Cerro Paranal Observatory. The main advantages of this instrument are a wide field-of-view of 1 squared arcmin, with a superb spatial resolution of 0.2" $\times$ 0.2" per spaxel, and a mean spectral resolution $R\sim 3000$ in the visible range 4650-9300 \AA.

Most of the observations were performed under the All-weather MUse Supernova Integral-field Nearby Galaxies (AMUSING; \citealt{Galbany2016}) survey, which is focused on the study of SN environments. Driven by different science goals, the sole criteria for a galaxy to be included in the AMUSING survey is to have hosted a classified SN of any type. For this reason, we have all types of galaxies, by morphology, mass or position in the sky in our sample. To obtain a good spatial resolution of the galaxies, we limited our sample to objects within a redshift ranging from 0.0005 to 0.076 (13.8 to 337.3 Mpc in distance). Therefore, the resolution in terms of physical scale goes from 13.3 to 327 parsec/pixel.

%In order to apply the reconstruction of the star formation history and the statistical analysis describe in sections~\ref{sec_ReconstrucSFH} and~\ref{subsec_statistic} we need that the galaxies satisfy two criteria, to have enough spaxels or zones with a good signal to noise ratio (The order of 100 zones), and cover the entire galaxy (2 or 3 effective radius). Too small or far away galaxies and too big or close one are rejected to the sample.

Our final sample is formed by 116 SNe in 102 galaxies, with 73 SNe Ia, 28 SNe II, and 15 SNe Ib/c  \citep[that groups stripped-envelope SNe: 5 Ic, 5 IIb, 2 Ib, 2 Ic-BL, and 1 Ibc;][]{Williamson2019}. SNe IIn are explicitly excluded, because the nature of their progenitors is still debated \citep{Fox2015, Dwarkadas2011}.
The SN sample includes events from 1970 to 2018, and the discovery of these SNe was made by different telescopes and sky monitoring programs \citep[e.g.][among others]{LOSS, ASASSN}. %such as the All-Sky Automated Survey for Supernovae (ASASSN; \citealt{ASASSN}) or the Lick Observatory Supernova Search (LOSS;  \citealt{LOSS}).
Most of the data cubes were taken after the SN explosion, and some of the SN environments were observed and analysed in \citet{Kuncarayakti2018}.
The sample is not \change{homogeneous} and presents some selection biases, especially for CC SNe (which come from targeted and untargeted survey programmes), but is well suited to carry out the present work, since about 82\% of the host galaxies are spirals, of which 75\% have inclinations below 60 degrees and 91\% are located at a redshift below 0.04 (less than 190 Mpc away). Face-on galaxies avoid projection effects, that help us in the SFH reconstruction. While a low-z sample give us a better spatial resolution of the galaxies.

%-----------------------------------------------------------------------------------
\section{Estimation of the supernova rate}
\label{sec_SNrate}
%-----------------------------------------------------------------------------------

\subsection{Reconstruction of the star formation history}
\label{sec_ReconstrucSFH}
%-----------------------------------------------------------------------------------

The star formation history (SFH) of a galaxy region can be inferred from the observed spectrum of the spaxel. To obtain the past evolution of the galaxy, we need to decompose the spectrum into a sum of simple stellar populations (SSP; \citealt{FIT3D}): a group of stars that were born at a given time, so they have the same age and chemical composition. Assuming each SSP follows a certain IMF when their stars are born, it is possible to calculate its evolution with time.

We consider that every spaxel in a galaxy can be modelled by a succession of star formation episodes. This way, the spectrum of that region may be comparable with the spectrum of a certain linear combination of SSPs under the effect of the dust abortion, and the velocity dispersion. The observed spectrum also has the gas continuum and line emission.

In practice, the decomposition of a spectrum is complicated because of the great degeneracy between two SSPs that are close in age \citep[e.g.][]{CidFernandez2014}. They display very similar spectra, and there are a lot of different combinations of age, metallicity, and dust extinction that fit equally well a given galactic spectrum. Some of those solutions have no physical sense (e.g. negative coefficients) but, for every spectrum, there are also many physically plausible scenarios that are able to provide a good fit. The possible implications of this degeneration problem are discussed in more detail in section \ref{sec_discusion}.
In order to increase the signal to noise ratio (S/N) of the observations, which helps to provide more reliable SSP fitting results, we decided to add several low S/N spaxels in larger areas, a process called zonification.

We use the Pipe3D code \citep{PIPEI,PIPEII}, a fitting algorithm that decomposes an IFS data cube into several independent zones and separates the emission from the gas and the different stellar populations, providing the best-fitting coefficients for each SSP. We use the Granada-Miles SSP models template \citep{Miles2011, GonzalezDelgado2010} and assume a \citet{Salpeter1955} IMF.

Pipe3D starts by adjusting the non-linear parameters (dust extinction, velocity and velocity dispersion), adjusting the different parameters using Monte Carlo iterations and a few SSPs. Once the non-linear parameters have been adjusted, it seeks to adjust the resulting spectrum to a linear combination of the spectra of the full set of SSPs.

\subsection{Delay time distribution}
\label{subsec_DTD}
%-----------------------------------------------------------------------------------

Once we have the SFH of the cube of each galaxy, we only need a DTD function in order to compute the SNR spaxel by spaxel.
%Using the statistical analysis describe later in section \ref{subsec_statistic}, we will be able to measure if these DTD models reproduce the observed SN distribution.
In this section we present the four different DTD models used in this work, expressed in units of SN~yr$^{-1} \;$M$_{\odot}^{-1}$ in all cases.

We will first consider a Gaussian function as a basic DTD model, mostly appropriate for CC SNe.
We are interested in estimating the value of the variance $\sigma ^2$, that will give us a characteristic time between the star formation and the SN events, providing information about the life time of their progenitors and the time scale for SN feedback:
\begin{equation}
    {\rm DTD_{\rm cc}} \,(t) \; = \; \phi_{cc} \;\cdot\; e^{-t^2  / 2 \sigma^2 }.
\end{equation}

We will compare it with two DTD models recently proposed in \citet{zapartas2017}; a single stellar evolution model DTD$_{\rm za}$ (the classical CC SN progenitor scenario)
\begin{multline}
    {\rm DTD_{\rm Za}} \,(t) \; = \\
    \left\{ { \begin{array}{ll}
        0 & t < 3 Myr \\
        10^{-9} \cdot \{-2.83 + 8.70 \cdot log(t) - 2.07 \cdot log(t)^{2}\} / t  & 3 < t < 25 Myr \\
        10^{-8} \cdot \{-4.85 + 6.55 \cdot log(t) - 1.92 \cdot log(t)^{2}\} / t  & 25 < t < 48 Myr \\
        0 & t \geq 48 Myr
    \end{array}}
    \right \}
\end{multline}
and a model DTD$_{\rm zab}$ that takes into account the interaction in binary star systems:
\begin{multline}
    {\rm DTD_{\rm Zab}} \,(t) \; = \\
    \left\{ { \begin{array}{ll}
        0 & t < 3 Myr \\
        10^{-9} \cdot \{-2.65 + 7.51 \cdot log(t) - 0.98 \cdot log(t)^{2}\} / t  & 3 < t < 25 Myr \\
        10^{-8} \cdot \{-0.89 + 1.73 \cdot log(t) - 0.51 \cdot log(t)^{2}\} / t  & 25 < t < 48 Myr \\
        10^{-8} \cdot \{-3.46 - 2.98 \cdot log(t) + 0.65 \cdot log(t)^{2}\} / t  & 48 < t < 200 Myr \\
        0 & t \geq 200 Myr
    \end{array}}
    \right \}
\end{multline}
The latter can produce CC SN from progenitor stars less massive than 8 $M_{\odot}$ due to mass transfer mechanisms. These are the so-called late CC SNe, with \change{delta times} up to 200 Myr.
Here we assume that the SFH recovered with single stellar evolution SSPs is valid for a DTD model with binary interaction systems, i.e. that binary interaction is relevant for the SNR but not in the SFH reconstruction. The use of SSP templates that include binary systems \citep[e.g. BPASS;][]{BPASS2017} may become an improvement for future work.

In order to try to characterise SN Ia, we also consider a power law DTD model with a delay time:
\begin{equation}
    {\rm DTD_{\rm Ia}} \,(t) \; = \; \left\{ { \begin{array}{lcc}
                        0 & if & t < \Delta \\
                        \\ \phi_{Ia} \;\cdot\; t^{-\alpha } & if & \Delta \leq t 
                    \end{array}}
                    \right.
\end{equation}
In this case we are going to fit two parameters. $\Delta$ is the time delay between the star formation and the first SNe Ia in Myr, e.g. the lower turn-on age of the SN Ia DTD. This parameter is important to estimate the typical delay time of SNe Ia feedback. The $\alpha$ parameter is the index of the power law and defines the distribution of the SNe Ia at long term. It provides information on the expected SN Ia SNR in early type galaxies.

\begin{figure}
\centering
\includegraphics[width=0.5\textwidth]{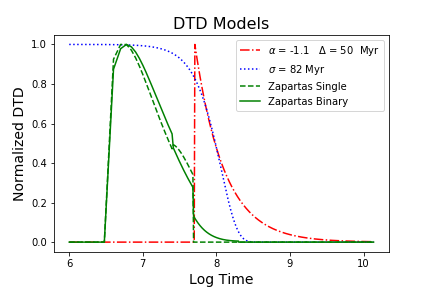}
\caption{Comparative representation of the different DTD models normalised to its maximum. In \change{dash-dotted line} a power law DTD with $\Delta = 50 Myr$ and $\alpha = -1.1$ is shown. In the \change{dotted line} a gaussian DTD with $\sigma = 82 Myr$ is shown. In solid line the Zappartas DTD for single stellar evolution, and in \change{dashed} line the model that includes binary interaction.}
\label{fig_DTD_Models}
\end{figure}

We represent these four DTD models in Figure \ref{fig_DTD_Models}.
The gaussian and the power-law DTDs have a normalisation parameter, $\phi_{cc}$ or $\phi_{Ia}$, typically given in number of SNe per stellar mass formed from the beginning of the \change{cosmos} ($N/M_{*}$). As will be explained below, for the purpose of this work we do not need to assume any normalisation factor of the DTDs, because we are looking for the relative distribution within each galaxy.

\subsection{Classical proxies}
\label{subsec_classical}
%-----------------------------------------------------------------------------------

\change{In addition to the above}, we consider two other parameters that have often been used as a proxy of the SNR: the emission in H$\alpha$ (proxy for young populations) and the total stellar mass (proxy for old populations), that are expected to correlate with the SNR of CC and SNe Ia, respectively.

Although these estimates are somewhat different in nature from our more elaborate procedure, they can be understood as a convolution of the SFH with a constant function (for the stellar mass) or a step-function DTD (for the last 30-40~Myr) in the case of H$\alpha$.
Most importantly, our approach does not depend on the specific method used to estimate the SNR $R(x,y)$ at every location within the galaxy; it would only evaluate whether it is statistically consistent with the observed distribution of SNe in the galaxy sample under study.

%-----------------------------------------------------------------------------------
\section{Statistical analysis}
\label{subsec_statistic}
%-----------------------------------------------------------------------------------

Once we have the SNR maps $R(x,y)$, we assume that the probability $p(x,y)$ of finding a SN in a certain place of a given galaxy is simply proportional to the local SNR.
%In this work the probability $(p)$ is assumed to be proportional only to the SNR.
More sophisticated analyses could be carried out, assuming some dependence of the detection probability with the dust extinction or the light contrast at different locations, in order to take into account observational biases (see the discussion in Section~\ref{sec_discusion}), but we think that these second-order corrections will only become important for larger galaxy samples.

The probability maps $p(x,y)$ can be normalised by imposing that the sum over the whole galaxy equals the number of known SNe in that galaxy.
In this way our results are independent of any normalisation factors in the theoretical DTD models, observational SNR tracers, or the precise proportionality constant in our simple $p(x,y) \propto R(x,y)$ prescription.

Starting from these individual probability maps, which tell us where a SN is most likely to be found within each galaxy, we test whether the observed locations are statistically compatible with being independent random events drawn from the proposed probability distribution.
Since we have just a few (usually only one) SN per galaxy, the only way to perform this test is to correlate the information of many different galaxies.
The methodology used here is the image pixel statistics using the normalised cumulative rank function proposed by \citet{James2006}.
The idea is to verify whether the distribution followed by the SNe is statistically consistent with the model predictions, i.e. that the number of SNe is proportional to the predicted probability $p$.
For this purpose, we calculate the cumulative probability
\begin{equation}
P_{i} \; = \frac {\sum\limits_{p_{j} \leq \; p_{i}} p_{j} } { \sum\limits_{j} p_{j} }
\end{equation}
at each point of our probability maps. Our definition of the cumulative probability is from lowest to highest $p$.
In this way the cumulative probability $P_i$ of a point $i$ is the sum over the values of $p_j$ lower than the local probability $p_i$ at that point.

In our normalisation, this cumulative probability $P$ goes from 0 to 1, being $P = 1$ the point with highest $p$ value.
If our probability models are correct, we expect a linear correlation between the fraction of SNe that explode in regions with $P_i < P$
\begin{equation}
F(P) \: = \; \frac{ n(P_i < P) }{ N },
\end{equation}
where $n\,(P_i < P)$ is the number of SNe that satisfy the condition and $N=116$ the total number of SNe in our sample, and the cumulative probability $P$.

For example, the value $P = 0.5$ divides the galaxies in two regions that have the same total probability of hosting a SN event.
Therefore, it is equally probable that a SN takes place in either region, and we would expect that half of the SNe have $P$ values lower than $0.5$.
This reasoning can be extended to any value of $P$, so we expect a simple linear behaviour of the form $F(P)=P$.

To measure the probability that our SNe follow the expected behaviour we use two different indicators: the Kolmogorov-Smirnov (K-S) test and the Anderson-Darling (A-D) test \citep{AndersonDarling1954}.
The K-S test  uses the maximum distance between a cumulative distribution and the theoretical model to calculate the probability that the two distributions are compatible. The A-D test works in a similar way, but instead of using the maximum distance between distributions, uses a quadratic weighted average distance. The Anderson-Darling test \change{gives more weight to} the tails of the distribution compared to the K-S statistics \citep{KSampleAD1987}

The two-sample K-S test is one of the most useful non-parametric methods for comparing two distributions, as it is sensitive to differences in both location and shape of the two samples. The K-S statistic value is the P value ($P_{v}$), the probability that the distributions do not reject the null hypothesis. Along the main body of the work we only use the K-S test, the A-D statistic results are very similar and are included in appendix \ref{Appendix_A}.

If the two distributions are drawn from the same parent population, when the size of the sample tends to infinity, the distance between them will tend to zero. However, our data samples are limited, so there is some uncertainty in the result. If the data sample is small, the uncertainty will be large. For this reason, in the case of the Ib/c SN we find higher $P_{v}$ values, since the large uncertainty allows the distribution to not reject the null hypothesis (values below 5\% or 10\%). Also, in this case it will be easier to find a parameters fit that was closer to a $P_{v} = 100\%$. With enough free parameters and a small data sample it is easier to maximise the $P_{v}$.

\begin{figure}
\centering
\includegraphics[trim= 0.9cm 0.5cm 1.5cm 0.5cm,clip=True,width=0.23\textwidth]{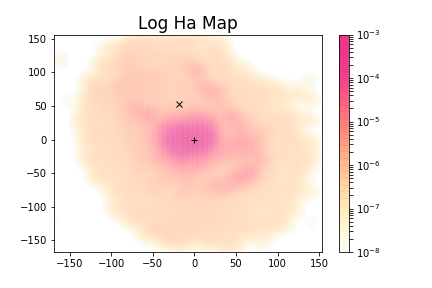}%
\includegraphics[trim= 0.9cm 0.5cm 1.5cm 0.5cm,clip=True,width=0.23\textwidth]{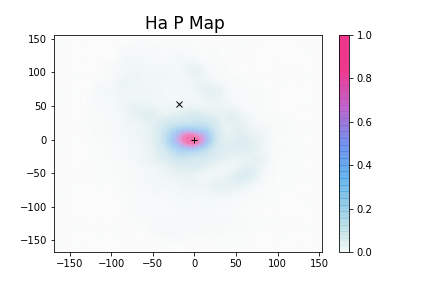}

\includegraphics[trim= 0.9cm 0.5cm 1.5cm 0.5cm,clip=True,width=0.23\textwidth]{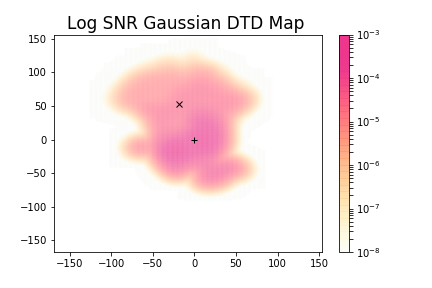}%
\includegraphics[trim= 0.9cm 0.5cm 1.5cm 0.5cm,clip=True,width=0.23\textwidth]{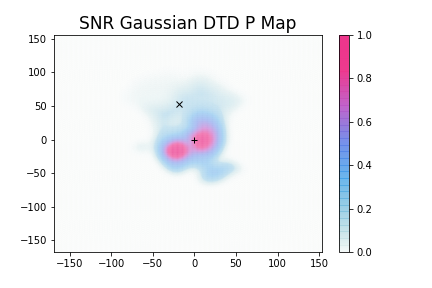}

\includegraphics[trim= 0.9cm 0.5cm 1.5cm 0.5cm,clip=True,width=0.23\textwidth]{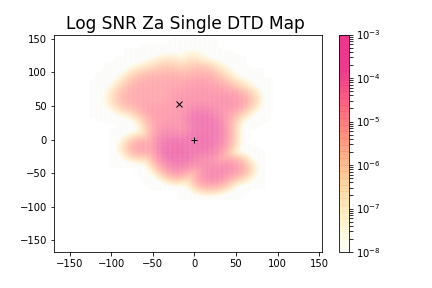}%
\includegraphics[trim= 0.9cm 0.5cm 1.5cm 0.5cm,clip=True,width=0.23\textwidth]{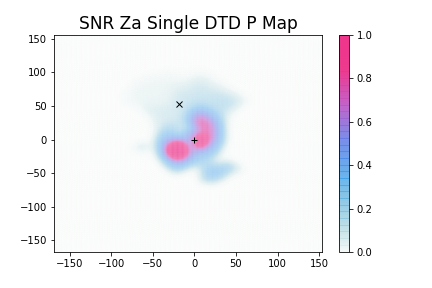}

\includegraphics[trim= 0.9cm 0.5cm 1.5cm 0.5cm,clip=True,width=0.23\textwidth]{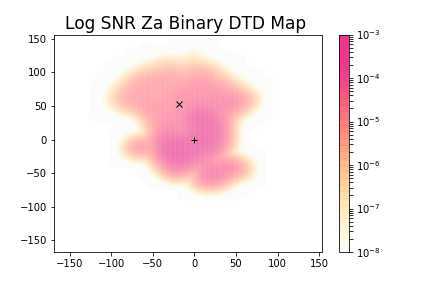}%
\includegraphics[trim= 0.9cm 0.5cm 1.5cm 0.5cm,clip=True,width=0.23\textwidth]{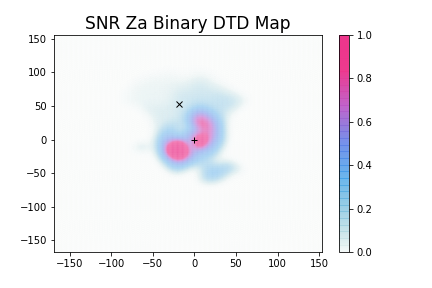}

\caption{We present the different probability maps for the SN 2006et as an example. In the first column we have the logarithmic representation of the normalised probability $p$. The right column presents the cumulative probability $P$. The tracers that appear are, in order: H$\alpha$, the Gaussian DTD SNR, the \change{Zapartas} single stellar evolution SNR and the \change{Zapartas} binary stellar evolution SNR. In all the maps the $X$ marks the position of the SN and the $+$ the center of the galaxy.}
\label{fig_Maps_CC}
\end{figure}

\begin{figure}
\centering
\includegraphics[trim= 0.9cm 0.5cm 1.5cm 0.5cm,clip=True,width=0.23\textwidth]{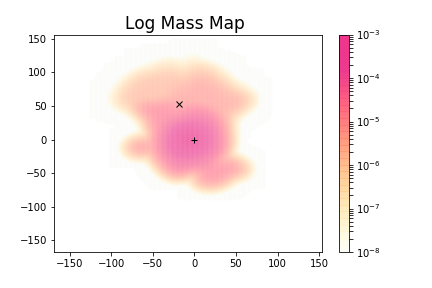}%
\includegraphics[trim= 0.9cm 0.5cm 1.5cm 0.5cm,clip=True,width=0.23\textwidth]{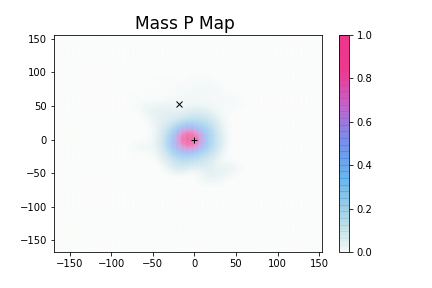}%

\includegraphics[trim= 0.9cm 0.5cm 1.5cm 0.5cm,clip=True,width=0.23\textwidth]{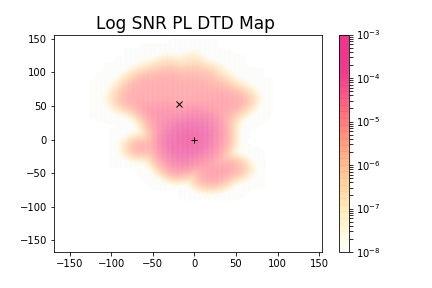}%
\includegraphics[trim= 0.9cm 0.5cm 1.5cm 0.5cm,clip=True,width=0.23\textwidth]{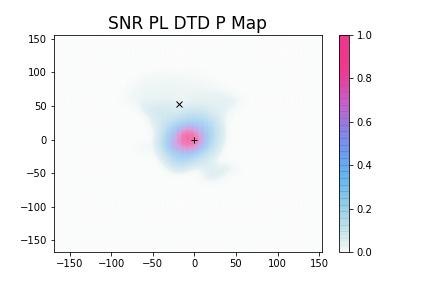}%

\caption{We present the different probability maps for the SN 2006et as an example. In the first column we have the logarithmic representation of the normalised probability $p$. The right column presents the cumulative probability $P$. The tracers that appear are, in order: the formed mass and the power law DTD SNR. In all the maps the $X$ marks the position of the SN and the $+$ the center of the galaxy.}
\label{fig_Maps_Ia}
\end{figure}

%-----------------------------------------------------------------------------------
\section{Results}
\label{sec_results}
%-----------------------------------------------------------------------------------

\subsection{Probability maps}
%-----------------------------------------------------------------------------------

We will consider six probability maps for each galaxy: four of them are expected to correlate with the CC SNR, and two of them correspond to the expected rate of Ia SN events.
For CC SNe, the H$\alpha$ map corresponds to the normalised surface brightness of the galaxy.
In the Gaussian DTD model we vary the $\sigma$ of the distribution from 10 to 700 Myr, and the Za and the Zab probability maps correspond to the SNR calculated from the \citet{zapartas2017} DTD models.
For SN Ia, we first compute the total mass formed in each spaxel according to the SFH reconstruction, which would roughly correlate with the SNR if the tardy scenario was their main production channel.
For the power law DTD models, we adjust the two free parameters $\Delta$ (the time delay, from 10 to 2000 Myr) and logarithmic slope $\alpha$ (from -0.5 to -2.1).

The best-fitting maps for CC and SNe Ia in the host galaxy of SN2006et \citep[NGC 232][]{LopezCoba2017} are illustrated in the left column of Figures~\ref{fig_Maps_CC} and~\ref{fig_Maps_Ia}, respectively. 
A spatial smoothing is applied to all our probability maps in order to minimise the effects of astrometric errors, as well as discreteness in the reconstruction of the SFH (Pipe3D divides the galaxy into zones to ensure a good signal to noise ratio), by means of a 2D convolution with a Gaussian with a standard deviation of 1.4 arcseconds (7 pixels).
The cumulative probability maps $P_{_{SN}}$ obtained by summation are represented on the right column panels.

\begin{figure}
\centering
\includegraphics[trim= 0.9cm 0cm 1.5cm 0.2cm,clip=True,width=\columnwidth]{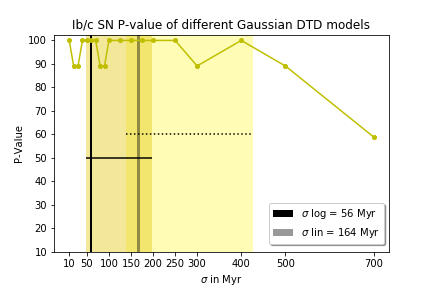}
\includegraphics[trim= 0.9cm 0cm 1.5cm 0.2cm,clip=True,width=\columnwidth]{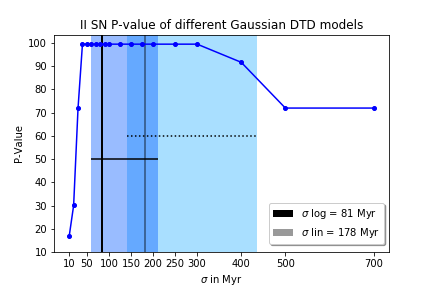}
\includegraphics[trim= 0.9cm 0cm 1.5cm 0.2cm,clip=True,width=\columnwidth]{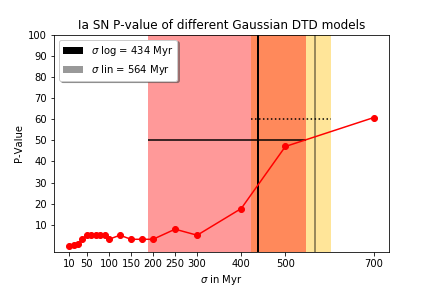}
\caption{Representation of probability value $P_{v}$ for the different $\sigma$ (standard deviation of the Gaussian DTD$_{\rm Ga}$) parameter. The top panel for the SNe Ib/c, the middle for the type II SNe and the bottom one for the Ia SNe. The \change{vertical line} represent the fiducial value for the two different priors, solid line for the logarithmic one and dotted line for the linear one.\change{The associated uncertainty, indicated by the solid bands, is estimated from the confidence interval that contains 25\% and 75\% of the probability ($\sigma_{25\%}$ and $\sigma_{75\%}$).}}
\label{fig_ccDTD}
\end{figure}

\begin{figure}
\centering
\includegraphics[trim= 0.9cm 0.3cm 2.0cm 0.2cm,clip=True,width=\columnwidth]{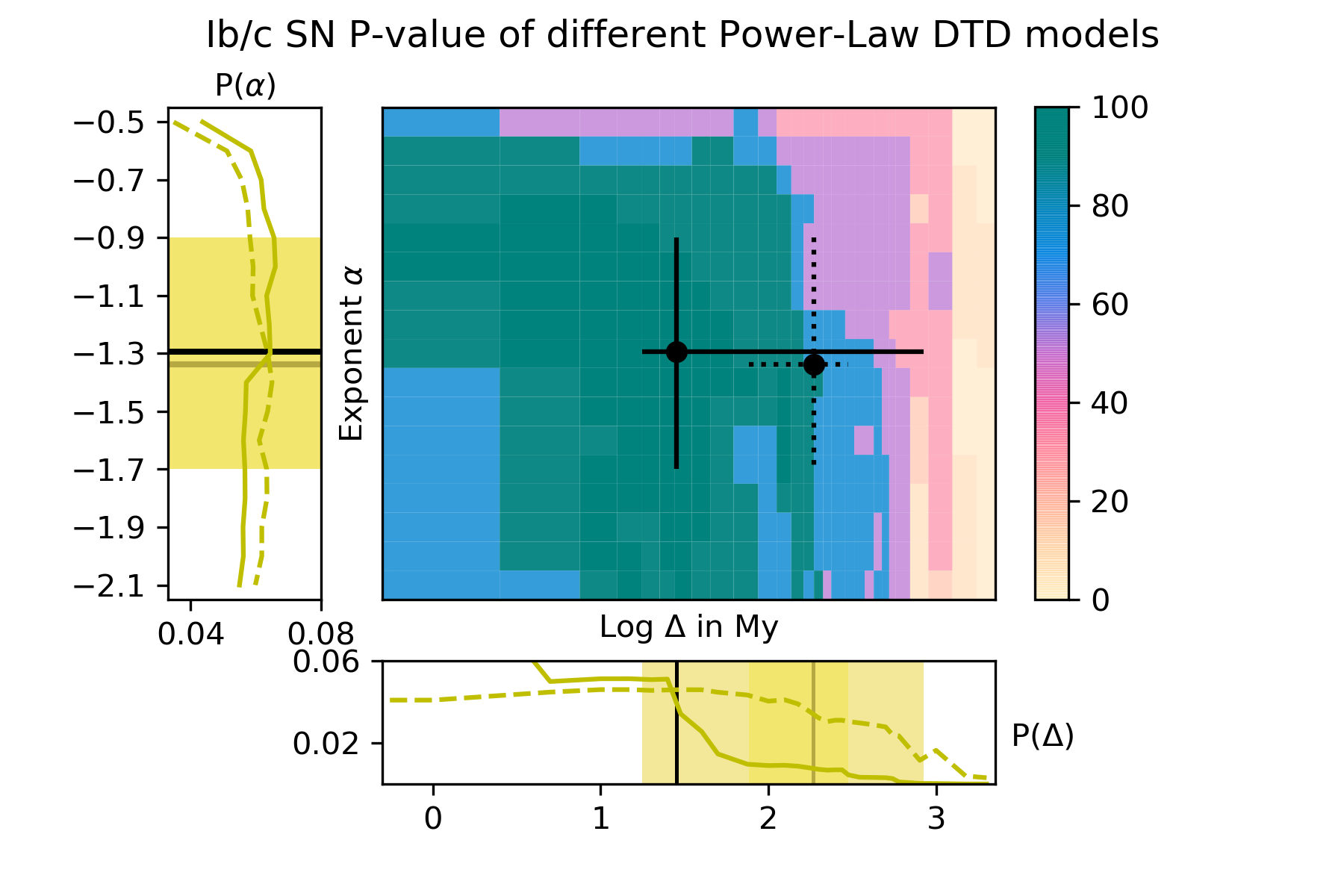}
\includegraphics[trim= 0.9cm 0.3cm 2.0cm 0.2cm,clip=True,width=\columnwidth]{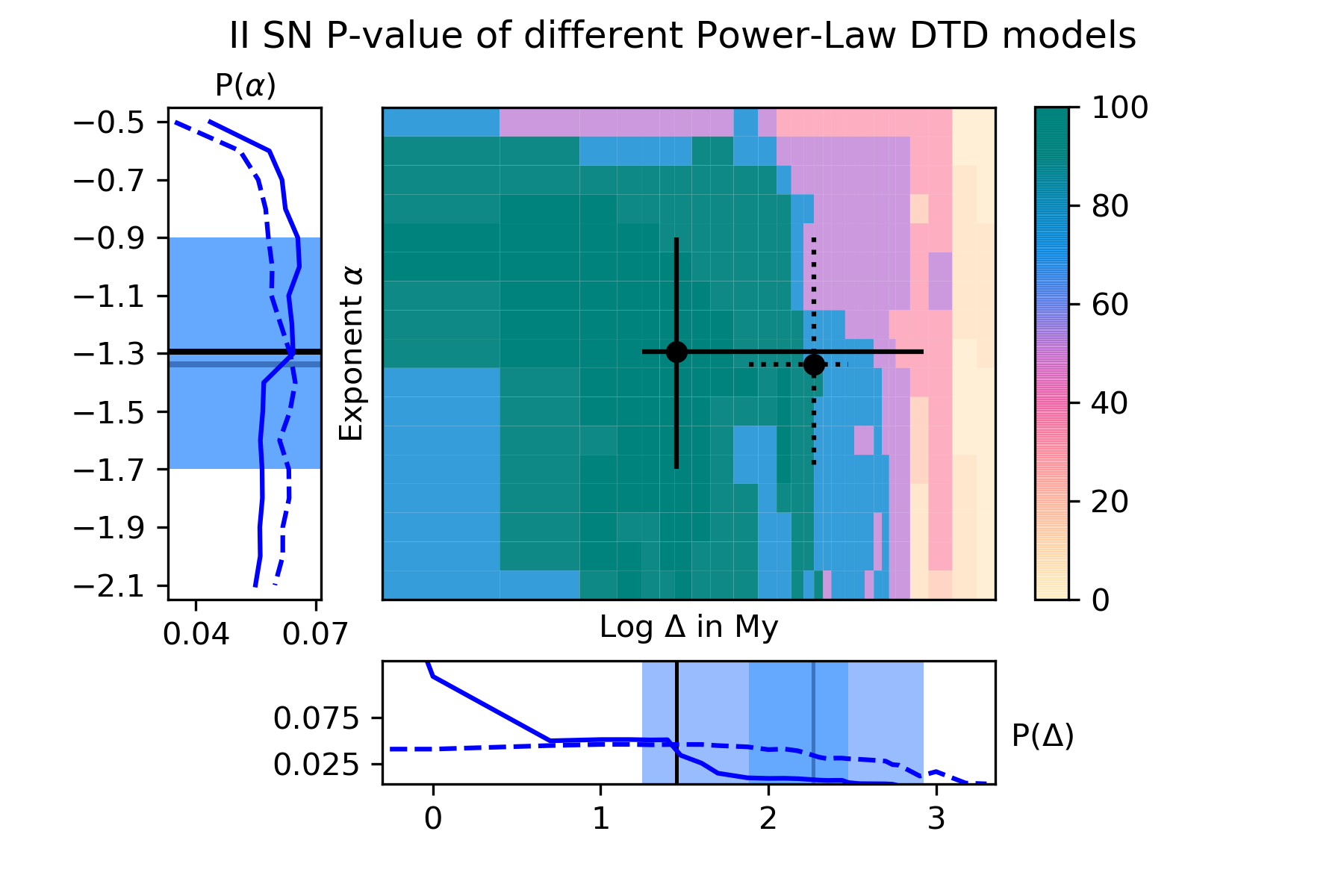}
\includegraphics[trim= 0.9cm 0.5cm 2.0cm 0.2cm,clip=True,width=\columnwidth]{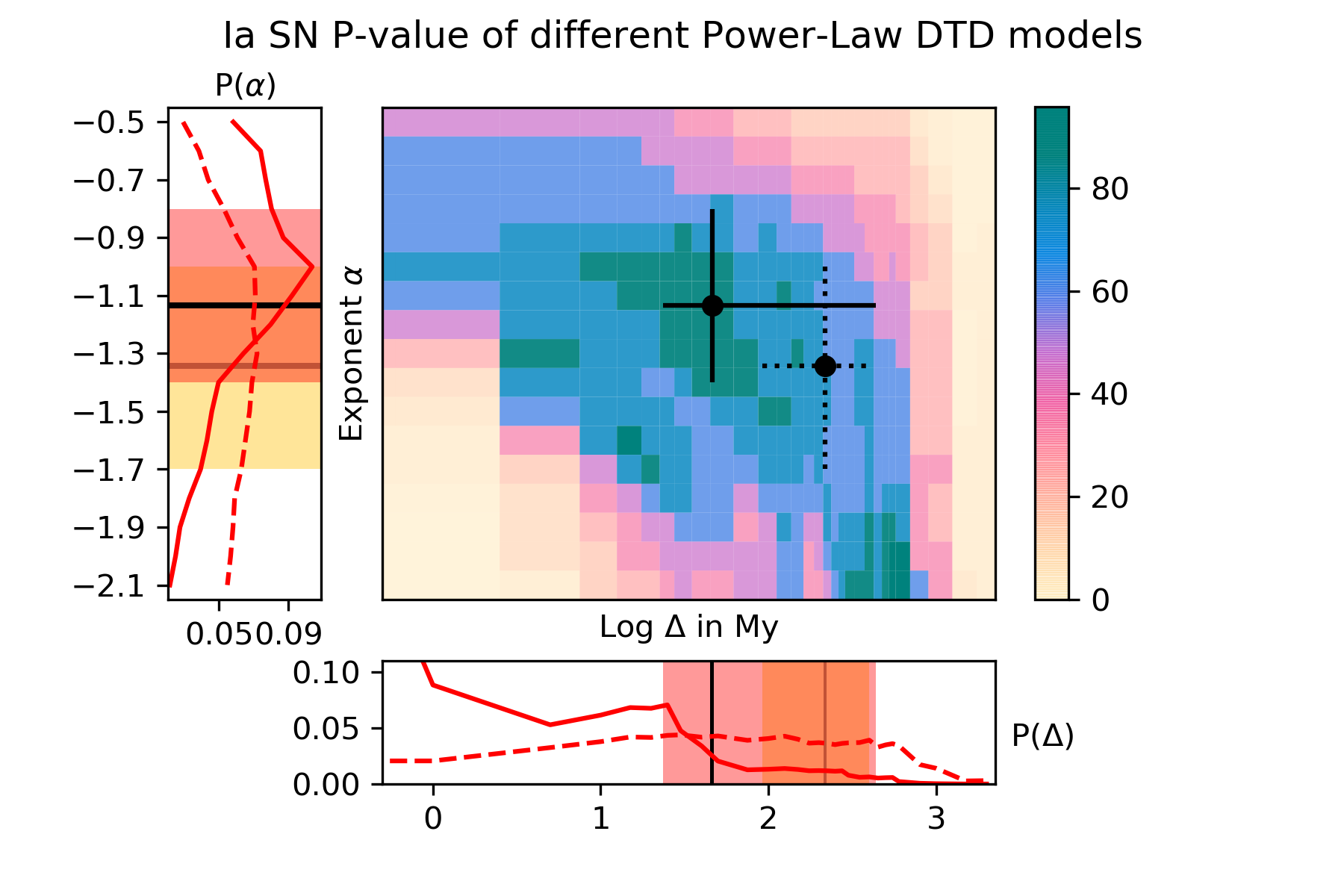}
\caption{Representation of probability value $P_{v}$ for the different $\alpha$ (Exponent of the power law) and $\Delta$ (Delay time) parameters. The first panel is for the SNe Ib/c, the second panel is for the type II SNe and the third one is for the Ia SNe. Adjacent panels represent the $P(\alpha)$ and $P(\Delta)$, the projected probability along these axes, the dashed line for a linear prior and the continuous line for the logarithmic one. The back dots represent our fiducial values and the lines the confidence interval for the two priors.}
\label{fig_IaDTD}
\end{figure}

\subsection{Fiducial parameters}
%-----------------------------------------------------------------------------------

In order to evaluate the quality of the fit in our parametric models, we split our sample according to SN type (Ib/c, II, and Ia).
Then, we compute the p-value with the K-S tests, comparing the cumulative fraction of galaxies $F(P_{SN}<P)$ with the linear distribution $F(P) = P$.
This would be the likelihood $\mathcal{L}$ of each model.
The results for the timescale $\sigma$ in the Gaussian DTD model, as well as the delay time and logarithmic slope in the delayed power law, are shown in Figures~\ref{fig_ccDTD} and~\ref{fig_IaDTD}, respectively.

\change{It can be clearly seen in Figure~\ref{fig_ccDTD}, the statistical distribution of SN Ib/c and II never drops below  10\%, so all models are consistent with the data. The best solution for SN Ib/c is found for $\sigma < 400$~Myr, and for SN II is found in the interval $40~Myr <\sigma < 400~Myr$.}
However, the distribution of SNIa is mostly inconsistent with this scenario.
The $p$-value increases with $\sigma$, but for $\sigma < 400$~Myr the $p_{v}$ is below 10\%, so the null hypothesis (the assumption that our DTD model does not correlate with the theoretical behaviour) can be rejected at the 90\% of confidence level.
For each class, we define our fiducial value of the timescale $\sigma$ according to Bayes' theorem:
\begin{equation}
    \langle \sigma \rangle = \frac{\int \sigma\, p(\sigma)\, \mathcal{L}(\sigma)\ \dd\sigma }{ \int p(\sigma)\, \mathcal{L}(\sigma)\ \dd\sigma }
\end{equation}
where $p(\sigma)$ denotes our prior probability distribution for this parameter.
We used two different prior distributions, uniform in linear and in logarithmic scale:
\begin{equation}
p(\sigma)_{\rm lin} = \frac{1}{\sigma_{\rm max}-\sigma_{\rm min}}
\end{equation}
\begin{equation}
p(\sigma)_{\rm log} = \frac{1}{\sigma \, ln (\sigma_{\rm max}/\sigma_{\rm min})}
\end{equation}
with $\sigma_{\rm max} = 700$~Myr and $\sigma_{\rm min} = 10$~Myr denote the adopted parameter range.

On the other hand, one can see in Figure~\ref{fig_IaDTD} that the parameters of the delayed power law models are poorly constrained when applied to the SN Ib/c and SN II observations.
The data are roughly consistent with any values $\Delta < 100$~Myr and $\alpha<-0.8$. Only the largest delta values ($\Delta \geq 1$~Gyr) reject the null hypothesis.
For SN Ia, the observed distribution can only be reproduced by \change{a turn-on time} of the order of tens of Myr, and logarithmic slopes around $\alpha\sim -1.1$.
The projected probability distributions of $\Delta$ and $\alpha$, marginalised by integrating along the other parameter according to Bayes' theorem
\begin{equation}
\noindent
P(\Delta) = \frac{ \int p(\Delta, \alpha)\, \mathcal{L}(\Delta, \alpha)\ \dd\alpha }{ \int p(\Delta, \alpha)\, \mathcal{L}(\Delta, \alpha)\ \dd\Delta\,\dd\alpha }
\end{equation}
\begin{equation}
P(\alpha) = \frac{ \int p(\Delta, \alpha)\, \mathcal{L}(\Delta, \alpha)\ \dd\Delta }{ \int p(\Delta, \alpha)\, \mathcal{L}(\Delta, \alpha)\ \dd\Delta\,\dd\alpha }
\end{equation}
are represented in the secondary panels of Figure~\ref{fig_IaDTD}, where lines and shaded regions indicate the expected value
\begin{equation}
\langle \Delta \rangle = \int \Delta \, P(\Delta)\ \dd\Delta ~~;~~
\langle \alpha \rangle = \int \alpha \, P(\alpha)\ \dd\alpha
\end{equation}
and the confidence intervals ($\Delta_{25\%}$ and $\Delta_{75\%}$) and ($\alpha_{25\%}$ and $\alpha_{75\%}$).

\begin{figure*}
\centering
\includegraphics[trim= 0.2cm 0cm 1.0cm 0cm,clip=True,width=0.48\linewidth]{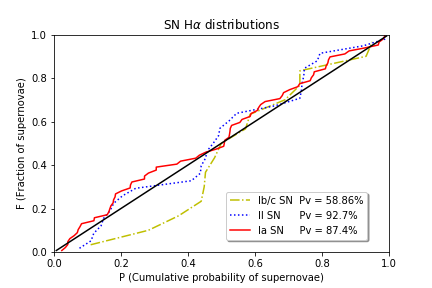}%
\includegraphics[trim= 0.2cm 0cm 1.0cm 0cm,clip=True,width=0.48\linewidth]{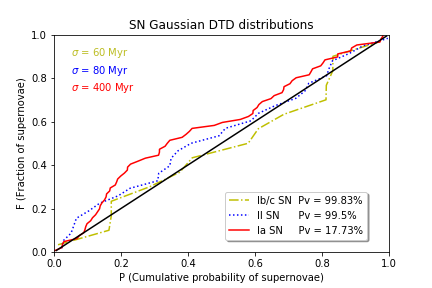}
\includegraphics[trim= 0.2cm 0cm 1.0cm 0cm,clip=True,width=0.48\linewidth]{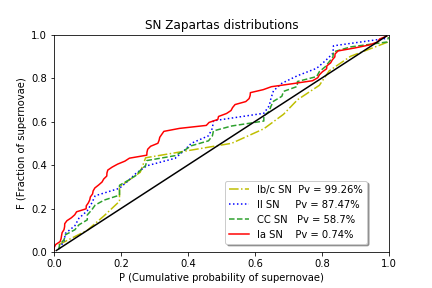}%
\includegraphics[trim= 0.2cm 0cm 1.0cm 0cm,clip=True,width=0.48\linewidth]{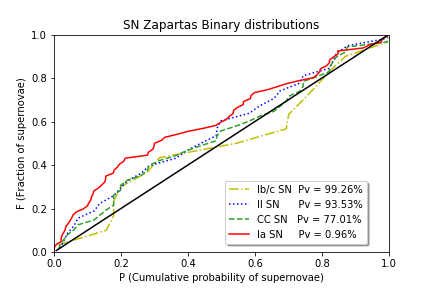}
\caption{Representation of the different fits for the cumulative distributions of SN probability $P_{_{SN}}$ against the fraction of galaxies $F(P)$. The top left panel is for the H$\alpha$ tracer, the top right panel is for the Gaussian DTD model, the bottom left panel is for the DTD$_{\rm za}$ model and the bottom right panel is for the DTD$_{\rm zab}$ model. SNe Ia, II, Ib/c and CC are indicated by \change{solid, dotted, dash-dotted and dashed} lines, respectively. The \change{solid straight} line shows the expected theoretical behaviour. In the key we show the $P_{v}$ of the cumulative distribution, the probability that the data points follow the theoretical behaviour.}
\label{fig_dist_cc}
\end{figure*}

\begin{figure*}
\centering
\includegraphics[trim= 0.2cm 0cm 1.0cm 0cm,clip=True,width=0.48\linewidth]{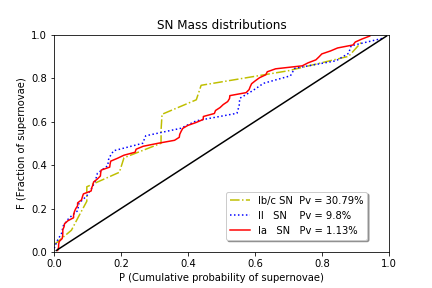}
\includegraphics[trim= 0.2cm 0cm 1.0cm 0cm,clip=True,width=0.48\linewidth]{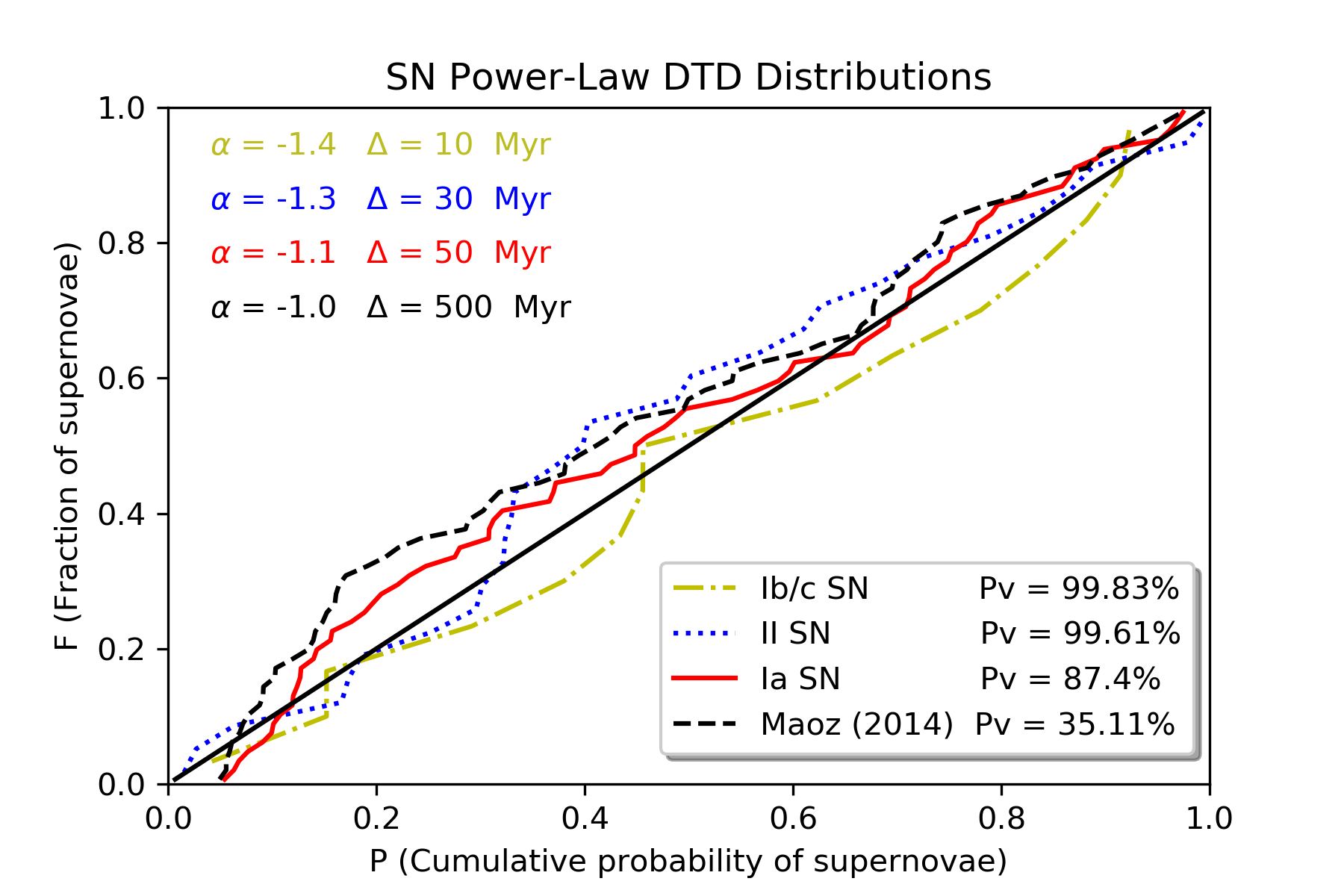}
\caption{Representation of the different fits for the cumulative distributions of SN probability $P_{_{SN}}$ against the fraction of galaxies $F(P)$. The left panel is for the mass formed and the right panel is for the power law DTD model. SNe Ia, II and Ib/c are indicated by \change{solid, dotted and dash-dotted} lines, respectively. The \change{solid straight} line shows the expected theoretical behaviour, and the dashed line in the right panel is the distribution for the $\alpha = 1$ power law DTD model from \citep{Maoz2014}. In the key we show the $P_{v}$ of the cumulative distribution, the probability that the data points follow the theoretical behaviour.}
\label{fig_dist_Ia}
\end{figure*}

\subsection{Cumulative distribution}
%-----------------------------------------------------------------------------------

Finally, we compare the cumulative distributions $F(P>P_{SN})$ obtained for each proxy of the SNR with the linear expectation.
We represent in Figures~\ref{fig_dist_cc} the tracers (H$\alpha$ emission, Gaussian DTD and Zapartas evolutionary models) that are expected to reproduce CC SNe, whereas those appropriate for SNe Ia are plotted in Figure~\ref{fig_dist_Ia}.
For the model parameters, we adopt the fiducial values inferred from logarithmic priors.

The correlation between the SN events with the H$\alpha$ emission is shown on the top-left panel in Figure~\ref{fig_dist_cc}.
In this case, we use the emission maps measured with Pipe3D. We obtain a very strong correlation for SNe II and Ib/c with a $P_{v} > 95\%$, as we expected. \change{We may expect} CC SN progenitors explode in less than 50 Myr, so we hope that they are very close to the HII regions where the progenitor stars were formed, which explains its relation with the H$\alpha$ flux \citep{Anderson2008}.
For SNe Ia, we find a fairly high $P_{v} = 86\%$, we do not reject the null hypothesis.
In principle, SNe Ia progenitors explode in a characteristic time larger than the scales probed by HII regions, so we do not expect such a strong correlation with the H$\alpha$ emission.
%This result could be explained by a 'prompt' population of SNe Ia, when the companion is a young star of high mass (in which case we would expect small $\Delta$ values, with a DTD maximum around 200-300 Myr), or it can be understood in terms of a strong correlation between the star formation rate at different epochs (i.e. a smooth star formation history).

The top-right panel in Figure~\ref{fig_dist_cc} shows the results of the Gaussian DTD model.
The p-values for the fiducial characteristic times $\sigma_{Ibc} = 50$~Myr and $\sigma_{II} = 70$~Myr are very close to 100~\%.
As shown in Figure~\ref{fig_ccDTD}, any values $\sigma_{Ibc} < 400$~Myr and $40~Myr < \sigma_{II} < 400$~Myr would be able to provide a reasonable fit.
Although, a larger sample would obviously make possible a more accurate estimation. Our current results are consistent with the theoretical expectations for both progenitor types.
For SNe Ia, we do not expect a good fit with the data.
So, it is thus not surprising that the fiducial model has a large $\sigma=400$~Myr and a p-value $P_{v} \simeq 20\%$. It rejects the null hypothesis for lower values of $\sigma$.

For DTD$_{\rm za}$ and DTD$_{\rm zab}$ (bottom left and right panel in figure~\ref{fig_dist_cc}), in this case we have also represented the distribution of CC SN with 43 SNe (28 Type II and 15 Ib/c) in order to increase the data sample and because the model used was made for CC SN all together. For the CC SN cumulative distribution we find higher $P_{v}$ in favour of the model that includes binary interaction ($P_{v}=77\%$), i.e. marginal evidence that, \change{at least, a population of late time core collapse supernovae with characteristic time from 80 to 200 Myr can fit the observations}. This result is in agreement with the previous characteristic time of $\sigma = 80 Myr$. In future works and with a larger data sample we could try to adjust different binary interaction channels for each of the CC SN subtypes.
The probability increase is also seen in the SNIb/c and SNII samples separately.
Once again, we find that SNe Ia do not fit at all to the model, with $P_{v}$ values less than 1\%. This result was expected since the \citet{zapartas2017} models are for CC SNe.

Regarding the models that are a priori more appropriate for SNIa (Figure~\ref{fig_dist_Ia}), we find that the galaxy mass (left panel) does not strongly correlate with any SNe type. This result was expected for the CC SNe since they  are assume to only depend on the recent star formation. SNe II have a very low $P_{v} = 10\%$, rejecting the null hypothesis, and SNe Ib/c only show a larger figure (30\%). \change{Taking into acount our understanding of the progenitors of SN Ibc, this effect can be explain by the small sample size}. A small sample size increases the uncertainty, making it more difficult for null hypothesis to be rejected.
For SNe Ia we expect a stronger correlation, but the results give the lowest $P_{v}=1\%$, rejecting the null hypothesis. We conclude that the total integrated stellar mass is a poor proxy of the Ia SNR. This is a clear result from our method.

For the power-law DTD model (right panel in Figure~\ref{fig_dist_Ia}), we obtain relatively high p-values for SN Ib/c and SN II for our fiducial models with $(\Delta, \alpha)= (\rm 10~Myr, -1.4)$ and $\rm (30~Myr, -1.3)$ respectively.
All in all, such combinations of short $\Delta$ and steep $\alpha$ are not so different in practice to a Gaussian DTD model, as they are dominated by the very recent star formation history.
For SNe Ia we find (see Figure~\ref{fig_IaDTD}) that the $P_{v}$ of the different models varies smoothly, and that we have a sufficiently large sample to constrain the model parameters.
We conclude that the power law DTD model is adequate to describe the physical mechanism that regulates these SN explosions, and we find a $P_{v} \simeq 88\%$ for our fiducial model with $\Delta = 50$~Myr and $\alpha = -1.1$. The $\alpha$ parameter is similar to the one found by \citet{Maoz2017, Graur2014}.
\change{We find a small $\Delta$ parameter, so it could point to an important `prompt' population of SNe Ia}. This does not contradict the results found in the analysts of the characteristic times of SNe, since the $\alpha$ parameter describes a slow temporal variation, which would explain the long expected lifetime for SN Ia progenitor systems, the `tardy' population. However, this `prompt' population could explain the correlation between SNe Ia and the H$\alpha$ flux. In addition, the value of $\alpha$ is in agreement with the classical value for SNe Ia, $\alpha = 1.0$ \citep{Maoz2014}. And the turn-on time for SNe Ia $\Delta$ is compatible with 40 Myr, the fiducial zero-order approximation - i.e. \citet{Georgy2013}, WDs can explode as SN Ia as soon as they form. (In both cases the difference we find is hardly statistically significant.)

%-----------------------------------------------------------------------------------
\section{Statistical analysis simulations}
\label{sec_simulation}
%-----------------------------------------------------------------------------------

In order to test the robustness of our new methodology, we perform a series of simulations. From a sample of IFS observed galaxies we create a synthetic SN sample that follows a specific DTD function at our choice. Once we have built that SN sample, we analyse it following the methodology described in the previous sections. In this way, we can study the possible biases of our methodology. We focus our analysis in two DTD scenarios, the differentiation between the Zapartas DTD with single stellar evolution and the one with binary evolution, and the correct recovery of the power law DTD  $\alpha$ and $\Delta$ parameterization.

The IFS galaxy sample selected for this exercise is the MaNGA-Pipe3D data products \citet{Sanchez2016b, Sanchez2018} sample, which contain 4862 galaxies. The MANGA galaxy sample has been analysed with the same Pipe3D code as the AMUSING main sample. Using the same SSP synthesis decomposition parameter and recovering the SFH of each spaxel under the same conditions as our main sample.

To create the synthetic sample of supernovae, we calculate the SNR of each spaxel convolving the SFH recovered with Pipe3D with a DTD function of our choice. Then, we make the assumption that the SNR of the spaxel is directly proportional to the probability of finding a SN in that spaxel. Using this probability distribution, we can generate a synthetic SN sample across the entire MaNGA-Pipe3D data products. In order to mimic the same selection criteria of our AMUSING observational SN sample (it is more likely to detect SNe in star forming galaxies through targeted monitoring campaigns), we create a larger SN sample (typically 5000 SNe) and only select a sub-sample of SNe from the most prolific galaxies.

Once we have the synthetic SN sample, we perform the same statistical analysis described in Section~\ref{subsec_statistic}. Thus, we obtain the probability $P_{v}$ that each DTD model fits the SN sample. Finally, we can compare the fiducial DTD parameters that we recover from this analysis with those that we plugged into our synthetic SN sample simulation.

For the Zapartas DTD scenario, we compare the $P_{v}$ of the models analysed with themselves (the DTD that generates the SN sample and the one used in the statistical analysis are the same) with respect to the crossover analysis. In the analysis with themselves, we expect higher $P_{v}$ for both Zapartas DTD models. This is a test for the synthetic SN sample construction. In the crossover case, we expect lower $P_{v}$, especially when we increase the SNe sample size. This would indicate that our methodology is capable of differentiating between the two DTD models.

\begin{figure}
\includegraphics[trim= 0.2cm 0cm 1.0cm 0cm,clip=True,width=\linewidth]{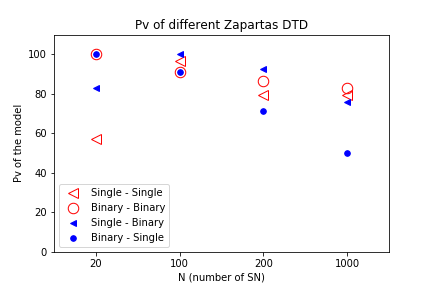}
\caption{We present the $P_{v}$ obtain from the simulated SNe sample analysis of the two Zapartas DTD models against the SNe Sample size N. The key shows the Zapartas DTD model used to built the SNe sample first, and the one used to obtain the represented $P_{v}$ in second place. Hollow markers are for the itself DTD correlation, and full ones for the crossover results.}
\label{fig_simulation_Za}
\end{figure}

The results can be seen in the figure~\ref{fig_simulation_Za}. We represent the different $P_{v}$ for a sample size N = 20, 100, 200 and 1000 SNe. It can be seen that all of the cases show similar $P_{v}$, especially for small sample size, and even in the N = 1000 case the $P_{v}$ are far away from the 10\% required to reject the null hypothesis. We conclude that these two DTD models are very similar and, therefore, we need very large sample sizes to achieve conclusive results using our methodology.

For the power law DTD scenario, we want to test the robustness of our methodology to find the underlying parametrization of a certain SN sample. For that reason, we construct a pool of simulated SN samples with different parametrizations. The $\alpha$ value takes the values [0.6, 1.0, 1.4, 1.8] and $\Delta$ is equal to [50, 100, 250, 500, 1000] Myrs. Thus, we perform our statistical analysis for these 20 SN samples, and compute its corresponding fiducial values.

\begin{figure*}
\centering
\includegraphics[width=0.5\linewidth]{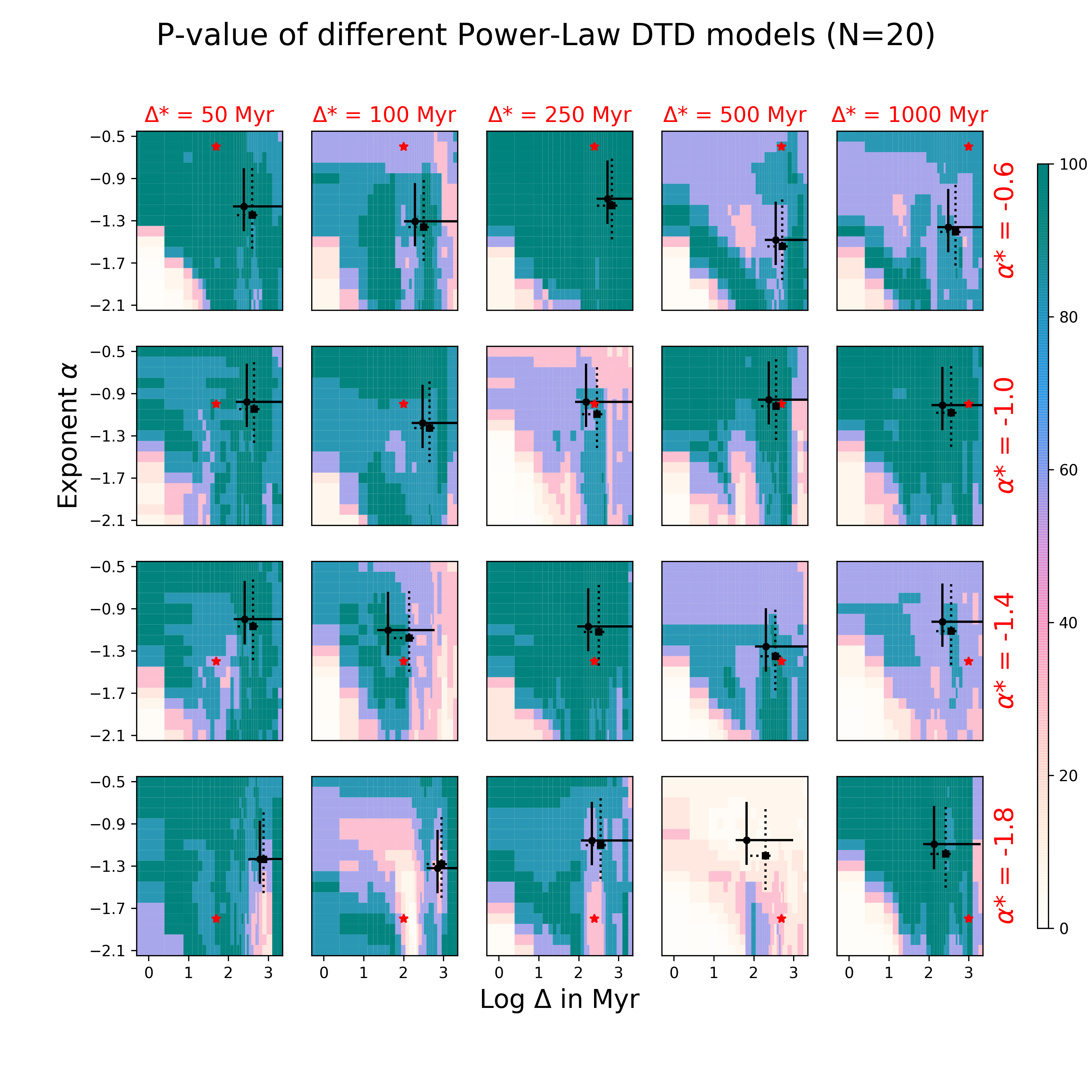}%
\includegraphics[width=0.5\linewidth]{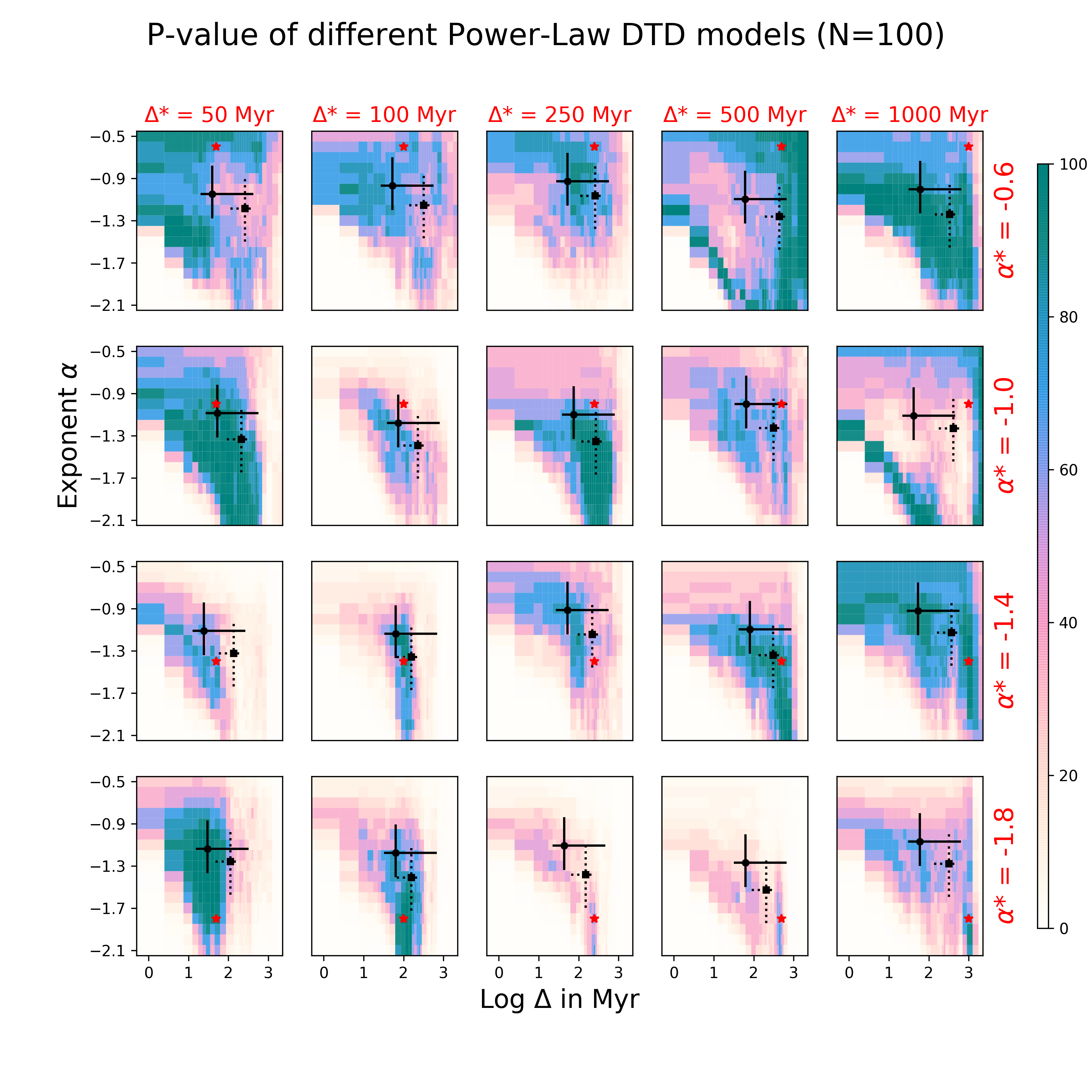}
\includegraphics[width=0.5\linewidth]{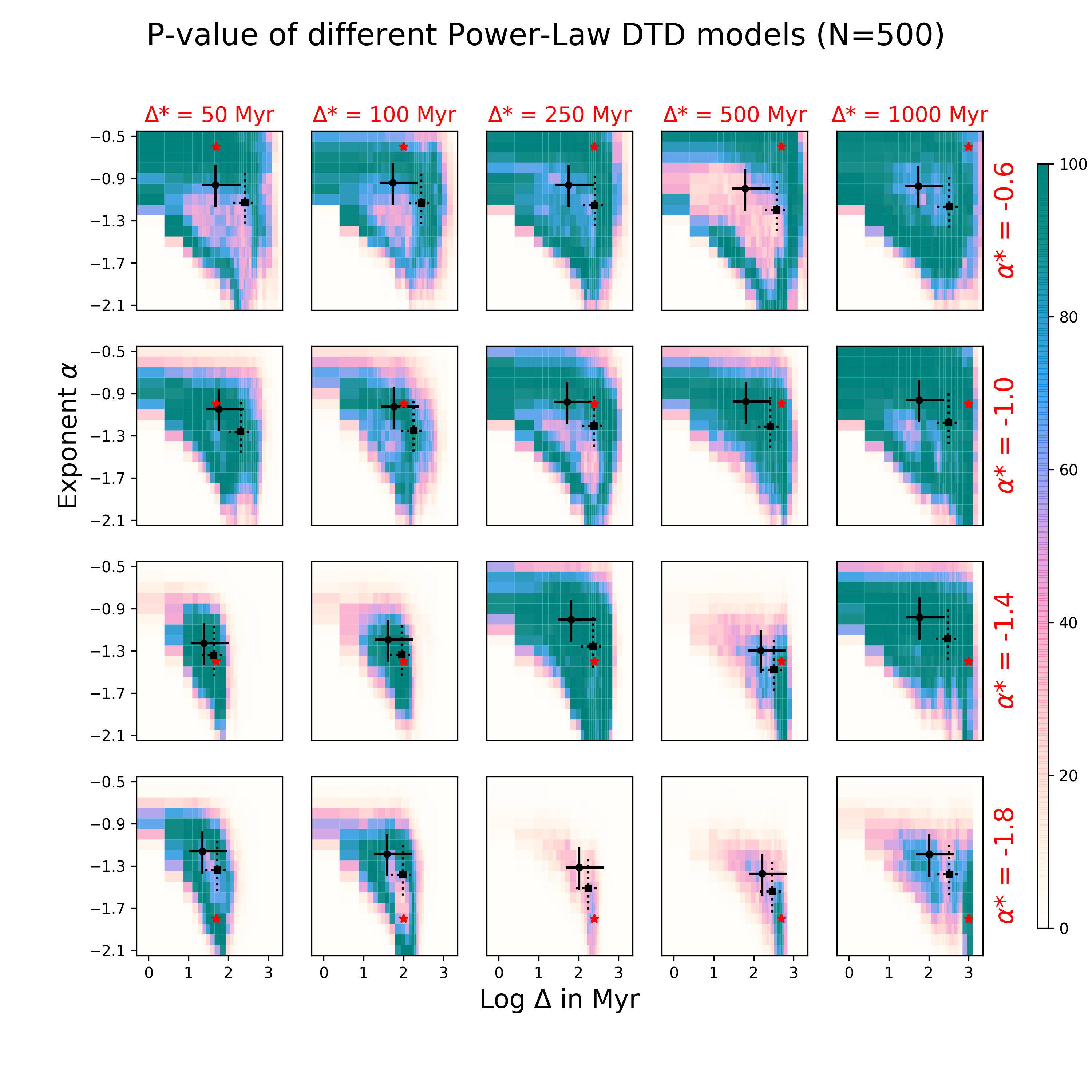}
\caption{The star values in the right and the top axis of the panels represent the parameter values use to construct each of the 20 simulated SN samples. The background plot represents the K-S $P_{v}$, the probability that a certain DTD parametrization fits the simulated SN sample. The $P_{v}$ values are shown in the right colour bar. The DTD model parametrization $\alpha$ and $\Delta$ are shown on the left and bottom axis of the panels. The star represents the parametrization used in the simulated SN sample construction. The black dots are the fiducial values find with the Bayesian statistics approach for the two prior used. The solid error bars are for the logarithmic prior, and the dashed line are for the linear one. The three different frames contain the panels for different SN sample sizes N = 20, 100 and 500.}
\label{fig_simulation_PLDTD}
\end{figure*}

The results can be seen in the figure~\ref{fig_simulation_PLDTD}. For each of the 20 synthetic SN samples we plot the different $P_{v}$ obtained, and compare the parameter values plugged into the SN sample (star values) with the recovered fiducial values (cross values). Different panels represent the different sample sizes N = [20, 100, 500].

For the case with N = 20, we find that all of the DTD parametrization are able to fits the SN sample, most of the $P_{v}$ are greater than 10\%. This can be explained due to the small sample size effect on the K-S tests performance. For the N = 100 and 500 case, we observe that our statistical analysis is being able to reject the null hypothesis ($P_{v}$ < 10\%) for some of the parametrizations. This is especially true for the short $\Delta$ and hard $\alpha$ combination of the bottom left corner. These DTD models yield SNR strongly dependent on the recent SFH, so it is expected that do not fit simulated SN samples produced by softer DTD functions.

In general terms, the N = 100 and especially N = 500 cases are able to obtain fiducial values that match the parametrization used in the simulated SN sample construction. So we can rely on our previous results and the robustness of the methodology. However, there are two caveats that we must highlight. On the one hand, when the simulated SN sample underlying parametrization are located near to the edges of the parameter space, the fiducial values obtained do not match the expected values. This effect is due to a prior bias in the parameter space, and it is especially important for the $\alpha$ = 0.6 and 1.8, and $\Delta$ = 1000 Myrs cases. Once the range of the parameter space is expanded, we are able to obtain the expected results. The other caveats to take into account is the existence of some degeneracies in the SNR calculation between different DTD parametrizations. This effect explains the ring shape that appears in some cases. A more consistent DTD normalisation should be able to break this degeneracy.

%-----------------------------------------------------------------------------------
\section{Discussion and conclusions}
\label{sec_discusion}
\label{sec_conclusions}
%-----------------------------------------------------------------------------------

In this work we constrain the DTD of different SN types using integral-field spectroscopy data of their host galaxies.
Here we summarise our main results and discuss some important caveats of the proposed approach.

Our sample includes SNe Ia, SNe II and stripped-envelope SNe Ib/c (Ib, Ic and IIb).
We do not consider individual types within the latter class because of the small sample size, but it would be interesting to do so in the future.
Possible classification errors are not taken into account, but we do not expect them to affect a significant fraction of the SNe, and therefore their statistical effect should be minimal.

In principle, biases in SN/galaxy selection should not pose an important shortcoming either, since our methodology does not depend on the fraction of the different SN types, and therefore it is not required that it is representative of the proportion in the real Universe.
SNe in distant galaxies are more difficult to detect, especially the intrinsically less luminous CC SNe, and targeted galaxy surveys, in contrast to untargeted samples, tend to focus on certain galaxy types and/or environments, as well as to avoid highly inclined objects.
These biases might have a some effect on the numerical results, but they are unlikely to significantly change the main results of our analysis, which is based on the spatial distribution of the SN explosions within each galaxy. %\change{This is supported by the fact that, we do not observe any important issue in the MANGA simulated sample analysis that is also affected by this SN/galaxy selection bias.}

A key assumption of the present work is that the probability $p$ of finding a SN in a certain spaxel of a galaxy is proportional to the SNR, neglecting the fact that some SNe may be undetected.
If the detection probability was constant within each galaxy, it would simply affect the number of SNe in our sample, and perhaps the fraction of different SNe types. However, the statistical analysis would be mostly unaffected.
Unfortunately, SNe taking place in the nuclear regions are more difficult to detect, due to the reduced light contrast with respect to the underlying stellar population and the potentially high dust extinction.
Based on \citet{Mattila2012} we estimate that we are losing about 15-20 \% of the SNe in the central region of the galaxies, and therefore we expect effects of this order on the significance of our results.
Given our relatively small sample size, our uncertainties are dominated by statistical errors, but in order to carry out a more precise analysis it is not only important to consider a larger galaxy sample, but also to introduce a prescription to account for the spatial variations of the SN detection probability in different spaxels.

With this caveat in mind, our statistical analysis method, based on the cumulative probability $P$ associated to each spaxel, may be used to compare the predictions of a given model with the observed distribution.
More precisely, we apply the Kolmogorov-Smirnov and Anderson-Darling tests to compute the p-value $P_{v}$ for different models, but other statistical descriptors may of course be used.
Based on these results, we use Bayesian inference to constrain the value of the free parameters of each model for different SNe types.
Once again, the relatively low number of SNe is the main limiting factor, and the results are somewhat dependent on the adopted priors.
We are fairly confident on the validity of our qualitative results and upper limits, but we caution the reader against an overinterpretation of the precise expectation values of each parameter, which should be regarded as a first order of magnitude estimate based on a new technique rather than an accurate measurement.

For instance, one can firmly rule out that the total stellar mass traces any SN type (only Ib/c SNe, due to their low number, would be marginally consistent from a strict statistical point of view, although the observed distribution is clearly far from uniform) or that the Zapartas DTDs, appropriate for CC SNe, correctly predict the Ia SNR.
Regarding the use of H$\alpha$ as a proxy of the SNR, processes different from star formation (like scattered light, turbulent collisions in the interstellar medium, supernova remnants, planetary nebulae, or active galactic nuclei) may contribute to the observed emission.
% -- Removed the reference about supernova remnants \citep{Vuceti2015}, because we should then include references for all processes.
Yet, our H$\alpha$ results show a strong correlation with the data, especially taking into account that no free parameter is associated to this tracer.
More statistics and a more elaborate model of the H$\alpha$ emission would be required in order to investigate this issue in deeper detail.

The main advantage of IFS data over other observational techniques is that it makes possible to reconstruct the SFH in each spaxel of the galaxy and estimate the SNR by assuming a certain DTD, thus constraining the nature of the SN progenitors.
Nonetheless, this is a rather degenerate inverse problem, as there are many different possible combinations of simple stellar population models that reach similar fits to the spectral data.
The spectral range of the MUSE instrument is not optimal in this respect, as it leaves out the blue and near infrared ranges, which are extremely helpful in discriminating the different SSPs.
In addition, possible incompleteness of the stellar libraries, errors in the star isochrones, and uncertainties in the adopted IMF or solar abundance \citep[see e.g.][]{Walcher2011, Conroy2013} may also introduce systematic biases in the SFH reconstruction.

The accuracy of the SFH reconstruction is certainly far from perfect. Indeed we do find that, especially for young stellar populations, the best-fitting solutions tend to favour certain SSP ages, perhaps because their spectra represent more closely those of typical star-forming regions.
Nevertheless, this methodology represents a significant improvement with respect to photometric techniques, and our results, obtained with Pipe3D and MUSE data, are already more precise than previous attempts based on the application of the VESPA code \citep[VErsatile SPectral Analysis][]{Tojeiro2007} to the SDSS spectra, whose SFHs, used by many authors in the present context \citep[e.g.][]{Brandt2010, MaozMannucci2012, MaozMannucci2012DTD, Graur2013, Heringer2017, Maoz2017, Heringer2019}, only contain a single age bin under 400~Myr.
In the future, the addition of ancillary multi-wavelength observations and the comparison of the results obtained with different fitting codes and (updated) stellar libraries will make possible to impose even tighter constraints on the recent star formation at each spatial location within the galaxies.

Even with the current sample size and temporal resolution in the SFH, our method provides useful insight on different theoretical DTD models.
Regarding CC SNe, both of the models proposed by \citet{zapartas2017} -- one assuming classic single stellar evolution and another including the different interaction channels in binary systems -- are statistically compatible with SNII and SNIb/c data.
\change{Grouping all CC SNe together, we find no evidence in favour of the DTD based on binary stellar evolution, with $P_{v}=77\%$, or the single stellar evolution model with $P_{v}=58\%$.
Neither of them is firmly rejected, though, more detailed studies could help to estimate the statistical significance of this late CC SN population} \citep{Kuncarayakti2018, Auchettl2019}.
\change{As we saw in the simulation section, much larger sample sizes would be needed in order to disentangle between these two prescriptions.
More importantly,}
the fact that a simple Gaussian model yields a better fit, with $P_v$ close to 100\%, for any value of $\sigma$ in the range between $\sim 50-60$ and $200-400$~Myr, hints that the data are more consistent with a DTD with a significant contribution on those time scales, even more than in DTD$_{\rm zab}$.
Alternatively, it is also possible that a certain fraction of the CC SNe progenitors are ejected from a binary system after the explosion of the companion \citep{Zapartas2019}, which would associate these events with an older stellar population.

We obtain very similar results for SNe Ib/c and SNeII, both in terms of p-value as well as confidence intervals for $\sigma$ in the Gaussian DTD model.
In particular, we are not able to discern whether the progenitors of Ib/c SNe are dominated by stars of more than 25 $M_{\odot}$.
Given the small number of direct progenitor detections \citep{Smartt2015} and the different mechanisms proposed to eject hydrogen envelopes, from  Wolf-Rayet winds to binary Roche-lobe overflows \citep{Nathan2011}, this is an extremely important question, but both a larger sample and a more accurate reconstruction of the SFH in the last 10~Myr are necessary in order address it.

For SNe Ia, our results have sufficient statistical significance to rule all models but the H$\alpha$ proxy and the delayed power-law DTD.
For the latter, we find the best fit for a logarithmic slope $\alpha \sim -1.1$ (with a 50\% confidence interval between $-0.8$ and $-1.4$) and delay time $\Delta \sim 50$ (between 15 and 150)~Myr.
This represents a significant improvement \change{in time resolution} with respect to previous work \citep[e.g.][]{Aubourg2008, Brandt2010} reporting evidence of a short turn-on point for Ia SNe, $\Delta < 500$~Myr.

In summary, the combination of normalised cumulative rank statistics and integral-field spectroscopic data represent a promising alternative to characterise the delay time distribution of different supernova types.

Based on the AMUSING sample, we obtain constraints on the DTD of type-Ia and CC SNe that are broadly consistent with previous studies based on other samples and methods.
More SNe would be required in order to increase the statistical significance of the results and provide tighter constraints, but such an increase in sample size should be accompanied by a more sophisticated modelling; the reconstruction of the recent SFH (on scales below 100 Myr) and the spatial dependence of the SN detection probability are two particularly important issues that require further consideration.
From an observational point of view, ancillary multi-wavelength data (especially in the ultraviolet and near infrared regimes) would be extremely helpful in order to increase the age resolution in the recent star formation history.

%------------------------------------------------------------------------
\section*{Acknowledgments}
%------------------------------------------------------------------------

This work has been supported by the Spanish \emph{Ministry of Economy and Competitiveness} (MINECO) through the MINECO-FEDER grant AYA2016-79724-C4-1-P.
L.G. was funded by the European Union's Horizon 2020 research and innovation programme under the Marie Sk\l{}odowska-Curie grant agreement No. 839090.
SFS is grateful for the support of a CONACYT grant CB-285080 and FC-2016-01-1916, and funding
from the PAPIIT-DGAPA-IA101217 (UNAM) project.
Based on observations made with ESO Telescopes at the Paranal Observatory 
(programmes 
60.A-9100,
60.A-9301,
60.A-9304,
60.A-9329,  %SV-AMUSING
094.B-0733
095.D-0091, %AMUSING
095.D-0172, %HK
095.B-0624,
096.D-0296, %AMUSING
096.D-0786, %JL
097.B-0165,
097.A-0366,
097.D-0408, %AMUSING
097.B-0640
098.D-0115, %AMUSING
099.D-0022, %AMUSING
0100.D-0341). %AMUSING
This project makes use of the MaNGA-Pipe3D dataproducts. We thank the IA-UNAM MaNGA team for creating this catalogue, and the Conacyt Project CB-285080 for supporting them.

\section*{Data Availability}

The data underlying this article will be shared on reasonable request to the corresponding author.

%%%%%%%%%%%%%%%%%%%% REFERENCES %%%%%%%%%%%%%%%%%%

\bibliographystyle{mnras}
\bibliography{biblio.bib}

\clearpage
%%%%%%%%%%%%%%%%% APPENDICES %%%%%%%%%%%%%%%%%%%%%

\appendix

\section{A-D statistic comparative}
\label{Appendix_A}

Here we present the main results obtained from the alternative statistical analysis using the A-D statistical parameter. We also include the previous results to compare in the next tables.

\begin{table}
\centering
Power-Law DTD
\begin{tabular}{lccccccc}
\hline
 & $\alpha_{25\%}$ & $\alpha$ & $\alpha_{75\%}$ & $\Delta_{25\%}$ & $\Delta$ & $\Delta_{75\%}$ & $P_{v}$ \\
\hline
$Ia_{KS}$ & -0.8 & -1.13 & -1.4 & 15 & 46 & 150 & 87.4\% \\
$Ia_{AD}$ & -0.9 & -1.22 & -1.5 & 15 & 46 & 175 & 87.4\% \\
\hline
$II_{KS}$ & -0.9 & -1.26 & -1.7 & 15 & 28 & 125 & 99.6\% \\
$II_{AD}$ & -0.9 & -1.44 & -1.7 & 10 & 24 & 75 & 99.8\% \\
\hline
$Ib/c_{KS}$ & -1.1 & -1.44 & -1.8 & 10 & 12 & 40 & 99.8\% \\
$Ib/c_{AD}$ & -1.0 & -1.40 & -1.8 & 10 & 11 & 30 & 99.8\% \\
\hline
\end{tabular}
\caption{We compare the results obtained through the K-S and A-D statistics for the parameters of the Power-Law DTD for each SN type. We show the $\alpha$ and $\Delta$ values that fit the model, as well as the confidence intervals at 25\% and 75\%. In the last column we indicate the $P_{v}$ of that adjustment.}
\end{table}
\begin{table} Gaussian DTD
\centering

\begin{tabular}{lcccc}
\hline
 & $\sigma_{25\%}$ & $\sigma$ & $\sigma_{75\%}$ & $P_{v}$ \\
\hline
$Ia_{KS}$ & 188 & 434 & 546 & 17.7\% \\
$Ia_{AD}$ & 154 & 470 & 502 & 34.4\% \\
\hline
$II_{KS}$ & 59 & 81 & 211 & 99.6\% \\
$II_{AD}$ & 51 & 67 & 219 & 99.5\% \\
\hline
$Ib/c_{KS}$ & 47 & 56 & 197 & 99.8\% \\
$Ib/c_{AD}$ & 47 & 55 & 212 & 99.8\% \\
\hline
\end{tabular}
\caption{We compare the results obtained through the K-S and A-D statistics for the parameters of the Gaussian DTD for each SN type. We show the $\sigma$ values that fit the model, as well as the confidence intervals at 25\% and 75\%. In the last column we indicate the $P_{v}$ of that adjustment.}
\end{table}
\begin{table}
\centering

\begin{tabular}{lcccc}
\hline
 & $H_{\alpha}$ & Mass & $DTD_{Za}$ & $DTD_{ZaB}$ \\
\hline
$Ia_{KS}$ & 87.4 & 1.1 & 3.2 & 5.1 \\
$Ia_{AD}$ & 73.6 & 0.2 & 1.7 & 2.6 \\
\hline
$II_{KS}$ & 92.7 & 9.8 & 92.7 & 92.7 \\
$II_{AD}$ & 85.9 & 9.5 & 54.6 & 69.8 \\
\hline
$Ib/c_{KS}$ & 58.8 & 30.8 & 88.9 & 99.8 \\
$Ib/c_{AD}$ & 54.4 & 19.1 & 99.8 & 99.8 \\
\hline
$CC_{KS}$ & - & - & 59.5 & 77.7 \\
$CC_{AD}$ & - & - & 57.9 & 76.2 \\
\hline
\end{tabular}
\caption{We compare the results obtained through the K-S and A-D statistics for the different DTD models. The $H_{\alpha}$ and mass as tracers of the SNR. As well as the two Zapartas DTD models consider in this work, where we include the value obtained for the entire sample of CC SNe. We show the $P_{v}$ we get for each SN type.}
\end{table}

\section{SN cumulative distribution}
\label{Appendix_B}
\begin{table*}
\begin{tabular}{lcccccccc}
\hline
   SN Names & SN type & P Ha &  P Mass &    P Ia &    P II &  P Ib/c &    P Za &   P Zab \\
            & & & & $\alpha=-1.1$ & $\sigma = 80$ & $\sigma = 50$ & & \\
            & & & & $\Delta=50$ & & & & \\
\hline
 ASASSN-13an &      Ia &  0.166790 &  0.214114 &  0.243218 &  0.276724 &  0.321923 &  0.300061 &  0.116094 \\
 ASASSN-13ar &      Ia &  0.176560 &  0.465018 &  0.425409 &  0.363850 &  0.288595 &  0.330685 &  0.138494 \\
 ASASSN-13cj &      Ia &  0.810587 &  0.797284 &  0.676667 &  0.662022 &  0.649561 &  0.666815 &  0.661863 \\
 ASASSN-13cp &      Ia &  0.183792 &  0.150432 &  0.199649 &  0.218518 &  0.225290 &  0.212504 &  0.087793 \\
 ASASSN-13cu &      Ia &  0.169977 &  0.202695 &  0.138889 &  0.140020 &  0.143530 &  0.142631 &  0.070562 \\
 ASASSN-13dd &      Ia &  0.617443 &  0.715837 &  0.200471 &  0.210252 &  0.211658 &  0.208866 &  0.632618 \\
 ASASSN-13dm &      Ia &  0.765077 &  0.893268 &  0.842663 &  0.841596 &  0.848205 &  0.848257 &  0.799899 \\
 ASASSN-14at &      II &  0.148410 &  0.101438 &  0.106677 &  0.114995 &  0.123078 &  0.120365 &  0.022611 \\
 ASASSN-14bf &      II &  0.495634 &  0.247709 &  0.513217 &  0.555628 &  0.642835 &  0.614712 &  0.129677 \\
 ASASSN-14co &      Ia &  0.282171 &  0.175525 &  0.110919 &  0.121403 &  0.115203 &  0.111918 &  0.111086 \\
 ASASSN-14cu &      Ia &  0.927760 &  0.956929 &  0.589892 &  0.473068 &  0.310316 &  0.400149 &  0.893017 \\
 ASASSN-14db &      Ia &  0.824447 &  0.775332 &  0.793141 &  0.799046 &  0.831244 &  0.822238 &  0.601441 \\
 ASASSN-14dz &      Ia &  0.703900 &  0.493821 &  0.346934 &  0.357552 &  0.374829 &  0.367184 &  0.360002 \\
 ASASSN-14hr &      Ia &  0.812900 &  0.919075 &  0.581131 &  0.576123 &  0.586600 &  0.598531 &  0.586071 \\
 ASASSN-14ig &      Ia &  0.434269 &  0.868373 &  0.812806 &  0.820333 &  0.815044 &  0.813720 &  0.758731 \\
 ASASSN-14jc &      Ia &  0.508664 &  0.308720 &  0.265012 &  0.271485 &  0.277695 &  0.278943 &  0.118214 \\
 ASASSN-14jg &      Ia &  0.561500 &  0.339581 &  0.734098 &  0.636759 &  0.313855 &  0.523215 &  0.026360 \\
 ASASSN-14kr &      Ia &  0.232600 &  0.182336 &  0.010028 &  0.006106 &  0.004881 &  0.006588 &  0.084765 \\
 ASASSN-14lp &      Ia &  0.777381 &  0.731257 &  0.000885 &  0.000132 &  0.000021 &  0.000249 &  0.613713 \\
 ASASSN-14lt &      Ia &  0.423556 &  0.120667 &  0.075437 &  0.079043 &  0.095973 &  0.087659 &  0.058449 \\
 ASASSN-14lu &      Ia &  0.270433 &  0.135259 &  0.122224 &  0.119648 &  0.116852 &  0.118904 &  0.058495 \\
 ASASSN-14lv &      Ia &  0.733930 &  0.687206 &  0.594550 &  0.593651 &  0.585779 &  0.588327 &  0.496022 \\
 ASASSN-14mf &      Ia &  0.983163 &  0.797507 &  0.779288 &  0.809524 &  0.838807 &  0.823354 &  0.525271 \\
 ASASSN-15aj &      Ia &  0.818050 &  0.834848 &  0.183648 &  0.184763 &  0.173729 &  0.177245 &  0.825977 \\
OGLE-2013-SN-015 &      Ia &  0.868844 &  0.466087 &  0.292371 &  0.286772 &  0.277079 &  0.281854 &  0.242346 \\
OGLE-2013-SN-123 &      Ia &  0.506970 &  0.107586 &  0.021577 &  0.022279 &  0.021165 &  0.021400 &  0.013842 \\
OGLE-2014-SN-019 &      Ia &  0.881259 &  0.964054 &  0.972266 &  0.976710 &  0.972478 &  0.979497 &  0.899905 \\
    LSQ12hxx &      Ia &  0.201701 &  0.067853 &  0.024183 &  0.026671 &  0.029861 &  0.028717 &  0.010737 \\
    LSQ14bbv &      Ia &  0.588780 &  0.085074 &  0.064020 &  0.062833 &  0.052021 &  0.052996 &  0.025875 \\
     SN1970A &      II &  0.452461 &  0.959808 &  0.852881 &  0.766382 &  0.626715 &  0.744184 &  0.880014 \\
     SN1985G &      II &  0.523166 &  0.553926 &  0.507931 &  0.497009 &  0.476162 &  0.487069 &  0.553065 \\
     SN1990Q &      II &  0.076142 &  0.301633 &  0.358703 &  0.389147 &  0.461496 &  0.438276 &  0.112887 \\
     SN1993R &      Ia &  0.970824 &  0.511588 &  0.965580 &  0.947548 &  0.898436 &  0.923701 &  0.374154 \\
     SN1994D &      Ia &  0.182390 &  0.616841 &  0.608603 &  0.566525 &  0.515307 &  0.547363 &  0.384557 \\
    SN1996aq &      Ic &  0.479538 &  0.223255 &  0.564331 &  0.659795 &  0.685598 &  0.693747 &  0.425013 \\
    SN1997dn &      II &  0.118837 &  0.307270 &  0.308250 &  0.292032 &  0.277193 &  0.283849 &  0.160339 \\
     SN1998X &      II &  0.794514 &  0.780258 &  0.740825 &  0.719313 &  0.726585 &  0.733416 &  0.546318 \\
    SN1998bw &   Ic-BL &  0.372404 &  0.021424 &  0.011674 &  0.012212 &  0.013827 &  0.013248 &  0.000264 \\
    SN1998dq &      Ia &  0.585708 &  0.524567 &  0.798140 &  0.826057 &  0.792696 &  0.785517 &  0.209051 \\

    \end{tabular}
\caption{Table of SN values for the cumulative probability $P$ of the different DTD models represented in Figure \ref{fig_dist_cc} and Figure \ref{fig_dist_Ia}.}
\end{table*}

\begin{table*}
\contcaption{}
\begin{tabular}{lcccccccc}
\hline
   SN Names & SN type & P Ha &  P Mass &    P Ia &    P II &  P Ib/c &    P Za &   P Zab \\
            & & & & $\alpha=-1.1$ & $\sigma = 80$ & $\sigma = 50$ & & \\
            & & & & $\Delta=50$ & & & & \\
\hline

    SN1999ee &      Ia &  0.158811 &  0.756217 &  0.312995 &  0.166735 &  0.122016 &  0.189394 &  0.424880 \\
    SN1999ex &      Ic &  0.109521 &  0.358974 &  0.285347 &  0.278336 &  0.257559 &  0.278535 &  0.322671 \\
    SN2000do &      Ia &  0.483130 &  0.691332 &  0.771282 &  0.782319 &  0.801623 &  0.795583 &  0.582623 \\
    SN2000ft &      II &  0.140568 &  0.597246 &  0.408082 &  0.398429 &  0.388033 &  0.395740 &  0.264787 \\
    SN2001da &      Ia &  0.611240 &  0.322245 &  0.392027 &  0.420304 &  0.533431 &  0.496980 &  0.167528 \\
    SN2001fv &      II &  0.741085 &  0.996905 &  0.932183 &  0.922507 &  0.784467 &  0.876276 &  0.981586 \\
     SN2002J &      Ic &  0.499728 &  0.380802 &  0.167970 &  0.155959 &  0.149718 &  0.156184 &  0.052001 \\
    SN2002hy &  Ib-pec &  0.448653 &  0.232222 &  0.834327 &  0.829942 &  0.822051 &  0.824221 &  0.195290 \\
    SN2003gh &      Ia &  0.681593 &  0.421791 &  0.514374 &  0.624910 &  0.582741 &  0.566012 &  0.166390 \\
    SN2004dg &      II &  0.734361 &  0.114663 &  0.150819 &  0.199805 &  0.195653 &  0.178201 &  0.098000 \\
    SN2004fd &      Ia &  0.685797 &  0.768139 &  0.126871 &  0.128807 &  0.133323 &  0.132929 &  0.518728 \\
    SN2004ff &     IIb &  0.159322 &  0.032179 &  0.139303 &  0.167715 &  0.209871 &  0.183837 &  0.021089 \\
    SN2004gs &      Ia &  0.968575 &  0.987632 &  0.650864 &  0.669505 &  0.712274 &  0.694070 &  0.919833 \\
     SN2005Z &      II &  0.178237 &  0.097565 &  0.027428 &  0.027621 &  0.029967 &  0.029283 &  0.044285 \\
    SN2005bg &      Ia &  0.819796 &  0.763618 &  0.978939 &  0.975873 &  0.966243 &  0.978018 &  0.741462 \\
    SN2005cu &      II &  0.491362 &  0.862038 &  0.819644 &  0.778520 &  0.810350 &  0.822093 &  0.625915 \\
    SN2005hc &      Ia &  0.768696 &  0.105840 &  0.069176 &  0.054893 &  0.038561 &  0.047913 &  0.030186 \\
    SN2005ir &      Ia &  0.775368 &  0.859953 &  0.737702 &  0.768200 &  0.772070 &  0.774374 &  0.847014 \\
    SN2005ku &      Ia &  0.227398 &  0.371673 &  0.222904 &  0.192689 &  0.157610 &  0.179511 &  0.190300 \\
    SN2005lu &      Ia &  0.366929 &  0.565185 &  0.443566 &  0.481047 &  0.489301 &  0.483218 &  0.479028 \\
    SN2005lw &      II &  0.204821 &  0.106472 &  0.052532 &  0.057301 &  0.062288 &  0.059814 &  0.027743 \\
    SN2005na &      Ia &  0.270258 &  0.682692 &  0.068950 &  0.068148 &  0.067470 &  0.067524 &  0.505493 \\
     SN2006D &      Ia &  0.041193 &  0.224606 &  0.062440 &  0.039421 &  0.022557 &  0.035281 &  0.070393 \\
    SN2006br &      Ia &  0.177894 &  0.741343 &  0.694355 &  0.620313 &  0.491382 &  0.586565 &  0.590278 \\
    SN2006cm &      Ia &  0.119211 &  0.072757 &  0.455700 &  0.484254 &  0.454930 &  0.444839 &  0.026294 \\
    SN2006cu &      II &  0.746130 &  0.561926 &  0.758641 &  0.844718 &  0.833197 &  0.833296 &  0.706224 \\
    SN2006ej &      Ia &  0.228426 &  0.057320 &  0.036369 &  0.036404 &  0.031690 &  0.033232 &  0.013602 \\
    SN2006et &      Ia &  0.041417 &  0.146684 &  0.153976 &  0.158072 &  0.158099 &  0.154236 &  0.041559 \\
    SN2006fo &     Ibc &  0.689975 &  0.472099 &  0.473389 &  0.351548 &  0.273640 &  0.315027 &  0.319660 \\
    SN2006hx &      Ia &  0.630525 &  0.599953 &  0.563311 &  0.579166 &  0.572805 &  0.565416 &  0.446206 \\
    SN2006my &      II &  0.410144 &  0.385005 &  0.190138 &  0.165410 &  0.112448 &  0.137237 &  0.177899 \\
    SN2006os &      Ia &  0.305296 &  0.158953 &  0.109616 &  0.106688 &  0.096418 &  0.099897 &  0.064618 \\
    SN2006ot &      Ia &  0.607097 &  0.132907 &  0.573642 &  0.558676 &  0.528714 &  0.535180 &  0.060893 \\
    SN2007ol &      Ia &  0.987964 &  0.976076 &  0.941603 &  0.941297 &  0.941531 &  0.941569 &  0.943058 \\
    SN2007rz &      Ic &  0.282154 &  0.307544 &  0.581534 &  0.578588 &  0.530763 &  0.544407 &  0.439479 \\
    SN2007so &      Ia &  0.525625 &  0.140773 &  0.021918 &  0.024879 &  0.024257 &  0.023333 &  0.082260 \\
     SN2008V &      II &  0.747775 &  0.738270 &  0.767946 &  0.629913 &  0.410194 &  0.484113 &  0.592841 \\
    SN2008ee &      Ia &  0.555157 &  0.682100 &  0.526351 &  0.537201 &  0.540790 &  0.536419 &  0.524200 \\
    SN2008ge &      Ia &  0.524254 &  0.833320 &  0.864357 &  0.855900 &  0.839554 &  0.848205 &  0.782096 \\
    SN2008ia &      Ia &  0.305494 &  0.204360 &  0.029327 &  0.021460 &  0.018284 &  0.021155 &  0.169370 \\
    
\hline
\end{tabular}

\end{table*}

\begin{table*}
\contcaption{}
\begin{tabular}{lcccccccc}
\hline
   SN Names & SN type & P Ha &  P Mass &    P Ia &    P II &  P Ib/c &    P Za &   P Zab \\
            & & & & $\alpha=-1.1$ & $\sigma = 80$ & $\sigma = 50$ & & \\
            & & & & $\Delta=50$ & & & & \\
\hline

     SN2009I &      Ia &  0.060258 &  0.304660 &  0.103644 &  0.098704 &  0.102625 &  0.103934 &  0.112538 \\
     SN2009N &      II &  0.439683 &  0.669840 &  0.525081 &  0.210456 &  0.099449 &  0.198947 &  0.074683 \\
     SN2009Y &      Ia &  0.023271 &  0.310361 &  0.000804 &  0.000415 &  0.000576 &  0.000719 &  0.266549 \\
    SN2009bb &   Ic-BL &  0.932156 &  0.257499 &  0.664113 &  0.721871 &  0.727210 &  0.701028 &  0.320041 \\
    SN2009dq &      II &  0.786525 &  0.812521 &  0.655905 &  0.637830 &  0.655518 &  0.657504 &  0.887327 \\
    SN2009iw &      Ia &  0.077660 &  0.137907 &  0.034361 &  0.046267 &  0.061990 &  0.057481 &  0.139379 \\
    SN2009jr &      Ia &  0.674638 &  0.466699 &  0.630809 &  0.645136 &  0.624850 &  0.637671 &  0.310773 \\
    SN2009md &      II &  0.167458 &  0.379894 &  0.400758 &  0.501145 &  0.680480 &  0.583292 &  0.378687 \\
     SN2010A &      Ia &  0.526566 &  0.283988 &  0.513104 &  0.598624 &  0.787844 &  0.730725 &  0.245151 \\
     SN2010F &      II &  0.110802 &  0.044524 &  0.022615 &  0.020692 &  0.016994 &  0.018728 &  0.009708 \\
    SN2010ev &      Ia &  0.453274 &  0.596182 &  0.222028 &  0.203323 &  0.190035 &  0.195742 &  0.375527 \\
    SN2010jr &     IIb &  0.081878 &  0.467696 &  0.520521 &  0.512049 &  0.480984 &  0.506403 &  0.072704 \\
    SN2011iy &      Ia &  0.069711 &  0.358630 &  0.140738 &  0.138761 &  0.108691 &  0.116724 &  0.379701 \\
    SN2011jm &      Ic &  0.950732 &  0.778275 &  0.956131 &  0.990418 &  0.999175 &  0.997217 &  0.699016 \\
     SN2012P &     IIb &  0.669120 &  0.115567 &  0.130745 &  0.175029 &  0.171431 &  0.153608 &  0.101830 \\
    SN2012au &      Ib &  0.572198 &  0.887246 &  0.822905 &  0.858263 &  0.885834 &  0.871066 &  0.919328 \\
    SN2012bu &      II &  0.437099 &  0.008791 &  0.055883 &  0.045092 &  0.034442 &  0.040484 &  0.000625 \\
    SN2012cu &      Ia &  0.530500 &  0.118880 &  0.063500 &  0.059252 &  0.032033 &  0.042054 &  0.031506 \\
    SN2012gm &      Ia &  0.082296 &  0.386049 &  0.326942 &  0.298837 &  0.279888 &  0.298630 &  0.129444 \\
    SN2012hd &      Ia &  0.728167 &  0.786028 &  0.995929 &  0.995352 &  0.993750 &  0.995453 &  0.791360 \\
    SN2012ho &      II &  0.660239 &  0.653797 &  0.839535 &  0.843847 &  0.834677 &  0.843726 &  0.163718 \\
    SN2013aj &      Ia &  0.072192 &  0.664278 &  0.107779 &  0.114845 &  0.106778 &  0.110285 &  0.573712 \\
    SN2013fz &      Ia &  0.378148 &  0.466699 &  0.605928 &  0.694672 &  0.814137 &  0.793689 &  0.399739 \\
    SN2014at &      Ia &  0.047492 &  0.137839 &  0.153637 &  0.140607 &  0.148461 &  0.151510 &  0.024346 \\
    SN2014cd &      Ia &  0.510009 &  0.554159 &  0.275515 &  0.292365 &  0.328164 &  0.321439 &  0.445649 \\
    SN2014cy &      II &  0.244265 &  0.312826 &  0.360056 &  0.371057 &  0.361947 &  0.363314 &  0.273973 \\
    SN2014dm &      Ia &  0.033906 &  0.253024 &  0.141061 &  0.143961 &  0.160626 &  0.154704 &  0.140950 \\
   SN2016aew &      Ia &  0.522488 &  0.818343 &  0.703986 &  0.598550 &  0.463202 &  0.514614 &  0.482393 \\
   SN2016bas &     IIb &  0.962775 &  0.641857 &  0.784397 &  0.726920 &  0.696880 &  0.716044 &  0.384510 \\
   SN2016gfk &      Ia &  0.120780 &  0.112838 &  0.216730 &  0.235994 &  0.315861 &  0.295262 &  0.015471 \\
   SN2016hmq &      II &  0.923808 &  0.926422 &  0.549793 &  0.493000 &  0.473100 &  0.498446 &  0.842246 \\
   SN2016iyz &      II &  0.457771 &  0.492101 &  0.284005 &  0.248520 &  0.245159 &  0.260594 &  0.425281 \\
   SN2016jga &      II &  0.738841 &  0.768548 &  0.745567 &  0.686243 &  0.648091 &  0.675512 &  0.719000 \\
   SN2018ezx &      Ia &  0.163901 &  0.794255 &  0.850840 &  0.838003 &  0.826298 &  0.834725 &  0.636871 \\
   SN2018kcw &      II &  0.996437 &  0.532236 &  0.996537 &  0.995161 &  0.995018 &  0.995664 &  0.554836 \\
   iPTF13bvn &      Ib &  0.734361 &  0.114663 &  0.150819 &  0.199805 &  0.195653 &  0.178201 &  0.098000 \\  

\hline
\end{tabular}

\end{table*}

%%%%%%%%%%%%%%%%%%%%%%%%%%%%%%%%%%%%%%%%%%%%%%%%%%

% Don't change these lines
\bsp	% typesetting comment
\label{lastpage}
\end{document}